\documentclass[aps,prx,reprint,showpacs,twocolumn,superscriptaddress,eqsecnum]{revtex4-2}
\usepackage{xr}
\usepackage{amsmath}
\usepackage{amsfonts,amssymb,amsthm,amsxtra}
\usepackage[ruled,vlined]{algorithm2e}
\usepackage[]{graphicx}
\usepackage{grffile}
\usepackage{mathrsfs}
\usepackage{framed}
\usepackage{bbm}
\usepackage{bm}
\usepackage{braket}
\usepackage{xcolor}
\usepackage{mathtools}
\usepackage[bookmarks=false,colorlinks=true,urlcolor=blue,citecolor=blue,linkcolor=blue]{hyperref}
\usepackage[]{qcircuit}
\usepackage{hyperref}
\usepackage[all]{hypcap}
\usepackage{url}
\usepackage{booktabs}

\DeclarePairedDelimiter{\floor}{\lfloor}{\rfloor}

\begin{document}

\title{Variational Microcanonical Estimator}

\author{Kl\'{e}e Pollock}
\affiliation{Department of Physics and Astronomy, Iowa State University, Ames, Iowa 50011, USA}

\author{Peter P.~Orth}
\affiliation{Department of Physics and Astronomy, Iowa State University, Ames, Iowa 50011, USA}
\affiliation{Ames National Laboratory, Ames, Iowa 50011, USA}
\affiliation{Department of Physics, Saarland University, 66123 Saarbr\"ucken, Germany}

\author{Thomas Iadecola}
\email{iadecola@iastate.edu}
\affiliation{Department of Physics and Astronomy, Iowa State University, Ames, Iowa 50011, USA}
\affiliation{Ames National Laboratory, Ames, Iowa 50011, USA}

\begin{abstract}
We propose a variational quantum algorithm for estimating microcanonical expectation values in models obeying the eigenstate thermalization hypothesis. Using a relaxed criterion for convergence of the variational optimization loop, the algorithm generates weakly entangled superpositions of eigenstates at a given target energy density. An ensemble of these variational states is then used to estimate microcanonical averages of local operators, with an error whose dominant contribution decreases initially as a power law in the size of the ensemble and is ultimately limited by a small bias. We apply the algorithm to the one-dimensional mixed-field Ising model, where it converges for ansatz circuits of depth roughly linear in system size. The most accurate thermal estimates are produced for intermediate energy densities. In our error analysis, we find connections with recent works investigating the underpinnings of the eigenstate thermalization hypothesis. In particular, the failure of energy-basis matrix elements of local operators to behave as \textit{independent} random variables is a potential source of error that the algorithm can overcome by averaging over an ensemble of variational states.

\end{abstract}

\date{\today}

\maketitle

\section{Introduction}

Calculating the ground state and thermal equilibrium properties of large and complex quantum systems remains a central task in contemporary quantum physics. While for integrable systems analytical techniques can often solve the problem, in generic nonintegrable systems such methods do not apply. In the last two decades however, efficient numerical methods have been developed to calculate ground-state and thermal properties in settings where the target state is only modestly entangled. Tensor network (TN) methods exploit the locality of physical Hamiltonians, in particular their area-law entangled ground states \cite{hastings_area_2007}, to find efficient representations of the wavefunction via truncated matrix product states on classical hardware \cite{vidal_efficient_2003}. Additionally, these efficient representations can be extended to Gibbs states at finite temperature via matrix product operators \cite{kuwahara_improved_2021}. Examples of algorithms based on TNs include the minimally entangled typical thermal state (METTS) algorithm \cite{stoudenmire_minimally_2010} for estimating canonical averages, and an algorithm for estimating microcanonical averages using time evolving block decimation (TEBD) \cite{schrodi_density_2017}. In higher than one spatial dimension however, the TN contraction step becomes hard \cite{schuch_computational_2007}, so that classical algorithms may not be sufficient for the simulation of even weakly entangled quantum systems.

\begin{figure}\label{fig:high_level}
    \centering
    \setlength{\belowcaptionskip}{-5pt}
    \includegraphics[width = 0.40\textwidth]{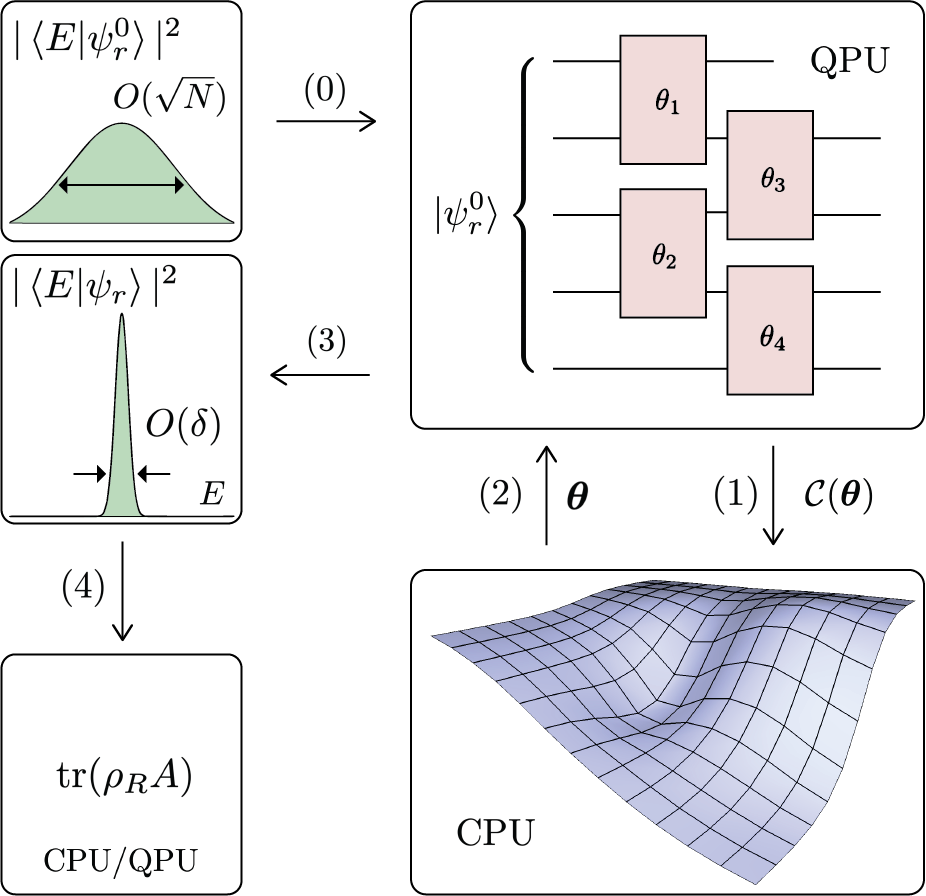}
    \caption{The VME algorithm. In step (0), the QPU is initialized in a random product state $\ket{\psi^0_r}$ ($r = 1, \ldots, R)$. The VQA repeats steps (1) and (2) that optimize the cost function $C(\boldsymbol{\theta})$ in Eq.~\eqref{eq:cost_function} to ``squeeze" the state onto a microcanonical window of size $\delta$ as shown in step (3). Steps (0-3) are repeated to produce a pseudo-random ensemble of states $\ket{\psi_r}$ which for large $N$ and $R$ can be used to approximate microcanonical averages of local operators $A$ as in step (4), where $\rho_R = \frac{1}{R}\sum_r \ket{\psi_r}\bra{\psi_r}$.}
\end{figure}

It has long been believed that quantum computers are the natural platform to simulate quantum systems~\cite{feynman_simulating_1982}, but to exploit their full power it is likely that deep quantum circuits and error correction will be required. Currently, we have noisy intermediate scale quantum (NISQ) devices that cannot yet implement deep circuits with high fidelity, but which can still demonstrate the potential for quantum computing in cases where low-depth circuits are sufficient \cite{preskill_quantum_2018}. There is thus a significant need to develop algorithms that can take advantage of these NISQ devices. 

Originating with the variational quantum eigensolver (VQE) \cite{peruzzo_variational_2014}, one class of algorithms that can potentially achieve this goal in some cases are the hybrid quantum-classical variational quantum algorithms (VQAs) \cite{tilly_variational_2022,cerezo_variational_2021,mcclean_theory_2016}, which employ a digital quantum computer aided by a classical optimizer. Although generic VQAs suffer from the well known barren plateau problem \cite{mcclean_barren_2018,larocca_diagnosing_2022,holmes_connecting_2022} which suggests unscalablility in full generality, there is evidence that VQAs can calculate the ground state of certain Hamiltonians using only polynomial quantum resources, e.g.~by using the Hamiltonian variational ansatz for the transverse field Ising model~\cite{wiersema_exploring_2020}. Recent works have also considered using VQAs to prepare Gibbs states using cost functions such as the relative entropy or relative free energy between the current state and target state \cite{verdon_quantum_2019}; strategies to overcome the costly evaluation of the entropic term have also been proposed \cite{foldager_noise-assisted_2022,selisko_extending_2022}. Other finite-temperature VQAs prepare thermofield-double (TFD) states, which require doubling the number of qubits in the physical system being simulated---for example the algorithm of Ref.~\cite{wu_variational_2019} can prepare the TFD state of free fermions efficiently at any inverse temperature. Alternative quantum algorithms for preparing thermal states include a quantum version of the minimally-entangled typical thermal states algorithm (QMETTS) that involves imaginary time evolution on quantum hardware~\cite{motta_determining_2020}, and an algorithm based on random quantum circuits \cite{shtanko_preparing_2023}.

In this work, we task a VQA with calculating microcanonical averages of local observables in
a one-dimensional (1D) nonintegrable spin model.
Our work is partially inspired by analog quantum simulation \cite{lu_algorithms_2021} and classical tensor network \cite{banuls_entanglement_2020} algorithms for estimating microcanonical properties. The algorithm takes advantage of the eigenstate thermalization hypothesis (ETH), in particular the ``diagonal" ETH which states that in a nonintegrable model the energy-basis diagonal matrix elements $\braket{E|A|E}$ of an observable $A$ approach a smooth function $A(E)$ in the thermodynamic limit \cite{deutsch_quantum_1991,srednicki_chaos_1994}.

The algorithm, which we call the variational microcanonical estimator (VME), works as follows (see Fig.~\ref{fig:high_level}). We initialize the QPU in a random product state [step (0)] $\ket{\psi_r^0}$, whose energy variance is typically extensive in $N$ (the number of sites) \cite{banuls_entanglement_2020}. Given a target energy $\lambda$ and microcanonical window size $\delta$, a classical optimizer is then tasked with minimizing the cost function
\begin{equation}
    \mathcal{C}(\bm{\theta})=\bra{\psi(\bm{\theta})}(H-\lambda)^2\ket{\psi(\bm{\theta})}
    \label{eq:cost_function}
\end{equation}
[steps (1) and (2)] originally proposed in \cite{peruzzo_variational_2014}. However, instead of trying to reach a local or global minimum, we stop the optimization as soon as $\text{Var}(H)=\braket{(H-\braket{H})^2} \leq \delta^2$. This produces states whose energy support is roughly limited to the microcanonical window of interest [step (3)], and the resulting variational states $\ket{\psi_r}$ are then used to compute the expectation of a local observable $A$ by averaging $\braket{\psi_r|A|\psi_r}$ over $R$ variational states [step (4)]. The ensemble average in step (4) enables a parametric reduction in the error and is essential to the algorithm's performance.

We benchmark the VME algorithm on a nonintegrable Ising chain by comparing its estimates for local observables to corresponding Gaussian microcanonical ensemble predictions obtained from exact diagonalization (ED). Using numerical evidence in combination with the phenomenology of ETH, we conjecture that for local operators $A$ and target energies $\lambda$ in the bulk of the spectrum, the absolute error in the VME algorithm scales as
\begin{equation}\label{eq:total_error_bound}
    \epsilon_R \simeq |c| + O(R^{-1/2}) + O(\delta/N) + O(\mathcal{D}^{-1/2}(\lambda)).
\end{equation}
Here, $\mathcal{D}(\lambda)$ is the density of states at the target energy $\lambda$, $\delta$ is the microcanonical window width, $N$ is the system size, and $c\ll 1$ is a small empirical constant whose magnitude depends on $A$ and other problem parameters. The last two terms in this formula are predicted by ETH and the first two terms we give a phenomenological argument for that we substantiate with numerical evidence.


We then generalize the problem to the reduced state of small subsystems of the chain and find numerically that when choosing $R = O(N^2)$ and for certain $\lambda$, the VME appears to approach the corresponding microcanonical state in the thermodynamic limit. The states prepared by the VME are consistent with area law entanglement for a fixed $N$, and require roughly linearly deep quantum circuits to prepare. We find that every random initial product state is able to converge, which we attribute to the fact that the algorithm does not seek global minima of the cost function. An additional distinction from other current VQAs for preparing mixed states is that we prepare pure states one at a time, thus avoiding storage of a large ensemble of pure quantum states in a quantum memory. The smallness of the trace distance when choosing $R= O(N^2)$ implies that the microcanonical ensemble, which involves at least one (via ETH) highly entangled (i.e. volume law) eigenstate is approximately indistinguishable \textit{by local operators} from a polynomially large ensemble of weakly entangled variational states.

The paper is organized as follows. In Sec.~\ref{section:motivation}, we introduce (i) the statement of ETH and (ii) a class of states which might be called microcanonical superposition states, which our converged variational states fall under. We then review related works attempting to use these states to estimate thermal averages and the relationship of this problem to ETH. In Sec.~\ref{section:var} we discuss how averaging over an ensemble of these microcanonical superposition states could significantly improve how well they can estimate microcanonical averages, and then we detail the VME algorithm which can produce these states. Finally in Sec.~\ref{section:numerical} we present the numerical results for the form of the variational ensemble, the error in the algorithm for various local operators, the observable independent trace distance, and finally the quantum resources like circuit depth and entanglement.

\section{Motivation}\label{section:motivation}

\subsection{Eigenstate Thermalization Hypothesis}

\begin{figure}\label{fig:eth}
    \centering
    \includegraphics{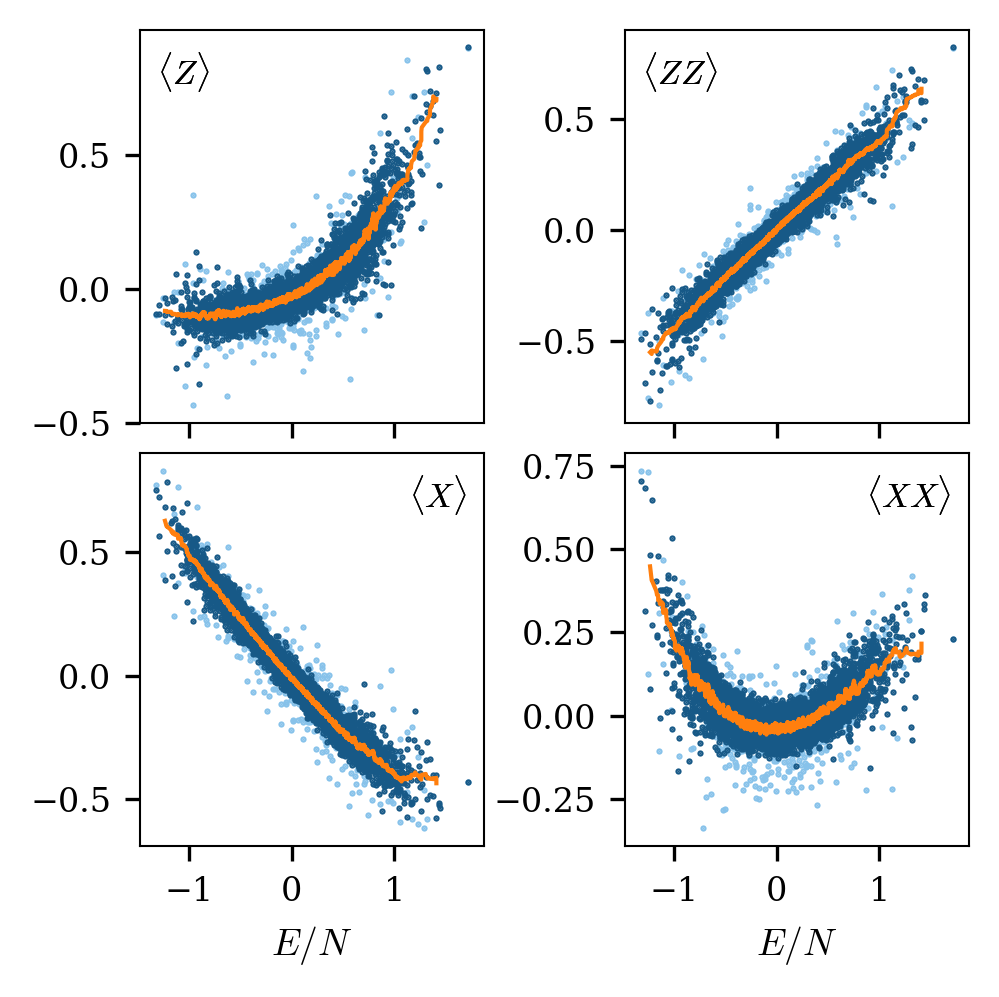}
    \setlength{\abovecaptionskip}{-5pt}
    \setlength{\belowcaptionskip}{-5pt}
    \caption{The energy-basis diagonal matrix elements $\braket{E|A|E}$ of various local observables $A$ acting in the middle of the chain, plotted against energy density in the nonintegrable 1D mixed-field Ising model, Eq.~\eqref{eq:mfim} with parameters $J = 1$, $h_x = -1.05$, and $h_z = -0.5$. Lighter blue colored points are for system size $N=9$ and darker blue points are for $N=13$. Orange curves are coarse grained versions of the $N=13$ scatter plots which define the ``smooth" function $A(E)$ in the thermodynamic limit.}
\end{figure}

Here we review relevant aspects of the ETH and some recent works which attempt to exploit it to estimate thermal averages. We assume a nonintegrable (i.e. chaotic) Hamiltonian $H$ which has a non-degenerate energy spectrum so that its eigenstates $\ket{E}$ are uniquely labeled by their energies $E$. Furthermore, we will assume that all operators and states of interest are real in the energy basis for simplicity. The variant of ETH we consider was formulated in Ref.~\cite{srednicki_approach_1999} and proposes that in a quantum chaotic system, the energy-basis matrix elements of observables have the form
\begin{equation}\label{eq:eth}
    \bra{E} A \ket{E'} = \delta_{EE'} A(\bar{E}) + \mathcal{D}^{-1/2}(\bar{E}) f(\bar{E},\omega) R_{EE'}
\end{equation}
where $\bar{E} = (E+E')/2$, $\omega = E-E'$, $\mathcal{D}(\bar{E})$ is the density of states at energy $\bar{E}$, $A(\bar{E})$ and $f(\bar{E},\omega)$ approach smooth functions in the thermodynamic limit, and $R_{EE'}$ are order-one fluctuations. Examples of such functions $A(\bar{E})$ are shown in Fig.~\ref{fig:eth} which demonstrates this for local spin operators in the 1D mixed-field Ising model (defined in Sec.~\ref{section:numerical}).

The ansatz \eqref{eq:eth} captures several features of such matrix elements that have been observed in numerical studies. Firstly, because the density of states is exponentially large in system size, the off-diagonal matrix elements are exponentially small. Secondly, the smooth function $A(E)$ is related to the statistical mechanical prediction for $\langle A \rangle$ at average energy $E$; this function will play a central role in our algorithm. Finally, the function $f(\bar{E},\omega)$ controls the approach to thermal equilibrium and is related to other spectral properties of the observable \cite{dalessio_quantum_2016}; this function figures less prominently in our analysis.

To see how $A(E)$ is related to a thermal average, consider for example a broadened microcanonical ensemble $\rho_{\lambda,\delta}$ centered on energy $E=\lambda$ and of width $O(\delta)$ which we will define more precisely at beginning of Sec.~\ref{section:numerical}. Under certain assumptions about the density of states of the model and away from $\lambda=0$ (which corresponds to infinite temperature), and assuming the ETH ansatz \eqref{eq:eth}, we have in the thermodynamic limit that (see Appendix~\ref{section:broadened} for details)
\begin{equation}\label{eq:smooth_eth}
      A(\lambda) = \langle A\rangle_{\text{mc}} + O(\delta^2/N) + O(\mathcal{D}^{-1/2}(\lambda))
\end{equation}
where $\langle A \rangle_{\rm mc} = \text{tr } \rho_{\lambda,\delta} A$. The ETH thus suggests that, if one could prepare even a single eigenstate $\ket{\lambda}$ of the Hamiltonian with energy $\lambda$, then one could accurately estimate thermal averages in sufficiently large systems. However for a nonintegrable Hamiltonian, a generic excited eigenstate is volume-law entangled, and thus cannot efficiently be prepared by classical algorithms nor by VQE-type algorithms \cite{zhang_adaptive_2021}. Thus, this feature of ETH does not appear practically useful, expect perhaps in the case of an error corrected quantum computer.

\subsection{Microcanonical Superpositions}
\label{subsec:microcanonical_superpositions}
An alternative approach to using exact eigenstates for computing thermal averages is using pure states of the form
\begin{equation}\label{eq:micro_sup}
   \ket{\psi} = \sum_E c_E \ket{E},
\end{equation}
where either $c_E$ are exactly zero outside the energy window defined by $|E-\lambda|\leq \delta$, or the states satisfy the weaker condition that $\braket{\psi| (H-\lambda)^2|\psi} = O(\delta^2)$. We refer to states of this type as ``microcanonical superposition states" and they have been studied in the context of thermal pure quantum (TPQ) states \cite{sugiura_thermal_2012}, the foundations of quantum statistical mechanics \cite{popescu_foundations_2006,goldstein_thermal_2015}, algorithms for analog quantum simulators \cite{lu_algorithms_2021}, and tensor network algorithms \cite{banuls_entanglement_2020}.

The practical reason for considering these states is that they appear to be significantly less entangled than exact eigenstates. In fact, there exist MPS-based numerical constructions of them such that the maximum entanglement entropy across any cut scales as $k/\delta + \text{log}_2\sqrt{N}$ for some constant $k$ \cite{banuls_entanglement_2020} and $N$ being the system size. Thus, by choosing $\delta = O(1/\text{log}_2N)$, such states can have only $O(\text{log}_2 N)$ entanglement, whereas a single excited eigenstate of a nonintegrable system is expected to have $O(N)$ entanglement. In this work, by choosing $\delta=O(N^{-1/2})$ (for the values of $N$ studied in this paper $N^{-1/2}\approx 1/\text{log}_2N)$ we find a VQA can generate these states using roughly linear circuit depth and which have area-law entanglement for fixed $N$. In Ref.~\cite{banuls_entanglement_2020} and in our findings it is clear that generically a smaller $\delta$ requires more computational effort.

It is known that if the coefficients $c_E$ are generic, and $\delta$ is sub-extensive in $N$, then when a state of the form \eqref{eq:micro_sup} is evolved under $H$, it approaches a state in which small subsystems are approximately thermal \cite{srednicki_approach_1999,rigol_thermalization_2008}. Given this fact, one may wonder if a relation like \eqref{eq:smooth_eth} holds with $A(\lambda)$ replaced by $\braket{\psi|A|\psi}$, just as it did for $\ket{\lambda}$. A key issue however is that although the off-diagonal elements of a generic operator are exponentially small, the quantity
\begin{equation}
    \braket{\psi|A|\psi} = \sum_E c_E^2 \braket{E|A|E} + \sum_{E\neq E'} c_E c_{E'} \braket{E'|A|E}
    \label{eq:psi_A_psi}
\end{equation}
involves summing exponentially many off-diagonal matrix elements, so long as $\delta$ itself is not exponentially small \cite{lu_algorithms_2021}. The reason that the long-time evolved state is locally thermal is that the off-diagonal terms become dephased just enough to counteract the exponentially large sum \cite{rigol_thermalization_2008}. As a result, the off-diagonal contribution scales as $O\bigl(\mathcal{D}^{-1/2}(\lambda)\bigr)$, like the off-diagonal matrix elements themselves. Without the additional pseudo-random phases $e^{i(E-E')t}$ appearing in the time evolved expectation value, for arbitrary $\delta$ there is no \textit{a priori} reason to expect $\braket{\psi|A|\psi}$ to closely approximate $\langle A \rangle_{\rm mc}$.

On the other hand, in discussions of ETH the off-diagonal fluctuations $R_{EE'}$ are usually stated to behave as random variables \cite{dalessio_quantum_2016}. If we take them to be actual \textit{independent} random variables, then the total off-diagonal contribution scales as a random walk and will therefore remain typically of the order $O(\mathcal{D}^{-1/2}(\lambda))$ as shown in Appendix~\ref{section:iid}. However, the validity of such an independence assumption on $R_{EE'}$ is known to depend on the energy scale $\delta$. Sometimes this scale is quoted as $\delta = O(N^{-2})$ which is the scale of $|\omega|$ below which $|f(0,\omega)|$ reaches a plateau, so that the ETH ansatz (at infinite temperature) becomes structureless and reduces to the random-matrix prediction \cite{dalessio_quantum_2016}. More recently, however, Refs.~\cite{richter_eigenstate_2020} and \cite{dymarsky_bound_2022} found numerical and analytical evidence that ``true" random matrix behavior with effectively independent matrix elements emerges only on the parametrically smaller scale, $\delta = O(N^{-3})$. 

Regardless of whether the matrix elements can be treated as independent random variables, it is in general an open question how $\delta$ must scale in order for $\braket{\psi|A|\psi}$ to converge to the thermal value in the thermodynamic limit. Ref.~\cite{banuls_entanglement_2020} argued that $\delta = O(1/\log_2{N})$ is sufficient for a slow convergence but Ref.~\cite{lu_algorithms_2021} argued that $O(N^{-1})$ is needed. Finally, in Ref.~\cite{dymarsky_new_2019} it is proposed that in a quantum chaotic system, for a fixed operator $A$ of interest, \emph{every} state of the form \eqref{eq:micro_sup} is thermal with a worst case error $x$ obeying the relation $\delta(x)=\text{poly}(x)$. However, for $\delta$ much larger than the random matrix theory scale $O(N^{-2})$ defined above~\cite{dalessio_quantum_2016}, the behavior (and in particular the $N$ scaling) of this polynomial was not completely settled in that work.

In summary, some source of randomness is needed to make the off-diagonal contribution small. In the theory of canonical typicality~\cite{popescu_foundations_2006,goldstein_thermal_2015}, it is the state coefficients; under time evolution it is the effectively random phases; if the window $\delta$ is small enough, it is the matrix elements themselves that are effectively random. In this work the source of randomness arises from averaging over an ensemble of variational states $\ket{\psi_r}$ that are prepared by starting from random product states $\ket{\psi_r^0}$. Since we will ultimately use shallow quantum circuits to prepare these variational states, we do not expect them to be typical states on the target microcanonical subspace, nor do we expect them to be typical states in the sense of TPQ states, as in both cases it is likely that deep circuits would be needed to approximate Haar random states \cite{brandao_local_2016,haferkamp_random_2022}. On the other hand, we will see that the states are ``random enough" for a certain dephasing mechanism to significantly reduce the off-diagonal contribution in the ensemble averaged version of Eq.~\eqref{eq:psi_A_psi}. Thus we will henceforth refer to the states as pseudo-random, reserving ``random" for Haar-random states.



\section{Variational Microcanonical Estimator}\label{section:var}

\subsection{Ensemble of microcanonical superpositions and error analysis}
\label{section:micro-sup}
As discussed in Sec.~\ref{section:motivation}, we do not necessarily expect a microcanonical superposition state of the form in Eq.~\eqref{eq:micro_sup} to closely approximate thermal values when the microcanonical window size $\delta$ is too large. Here we adapt known results from the theory of ETH to our problem, and then propose a heuristic mechanism which allows us to capture thermal expectation values using a large pseudo-random ensemble of microcanonical superposition states with ``large" (but still subextensive) $\delta$.

For a fixed target energy $\lambda$ and microcanonical window $\delta$, consider an ensemble of states $\{\ket{\psi_r}\}_{r=1}^R$, each of the form \eqref{eq:micro_sup}, and let $\rho_R = \frac{1}{R} \sum_r \ket{\psi_r}\bra{\psi_r}$ be their equal weight mixture. We discuss how to prepare these states using a variational algorithm in Sec.~\ref{section:algorithm}. In this section, we discuss the error between the actual microcanonical expectation value $\text{tr}(\rho_\text{mc} A)$ and its average in the ensemble $\rho_R$. Considering a local operator $A$, this error can be expressed as
\begin{equation}\label{eq:total_error}
    \epsilon_R = |\text{tr}(\rho_R A)-\text{tr}(\rho_\text{mc} A)|\,.
\end{equation}
Here, we have introduced a microcanonical density matrix $\rho_\text{mc}$. The microcanonical ensemble could be defined via the standard sharp microcanonical window, or via a smoothed Gaussian version thereof; we will ultimately compare our variational estimates to the latter in Sec.~\ref{section:numerical}. The results of this section do not depend on the exact form, but for now let us take $\rho_\text{mc}= P_W/n$ with $P_W$ a projector onto the microcanonical window $W$ defined by $|E-\lambda|\leq \delta$ and $n=\text{tr }P_W$ the number of states in the window. For our purposes, the error \eqref{eq:total_error} is best understood as a sum of two distinct types and we therefore decompose it via the triangle inequality as 
\begin{subequations}
\begin{equation}
    \epsilon_R \leq \epsilon^\text{diag}_R + \epsilon^\text{off}_R,
\end{equation}
where
\begin{align}
    \epsilon^\text{diag}_R &= |\text{tr}(\rho_\text{mc} A)-\langle A\rangle^\text{diag}_R| \label{eq:diag_error}\\
    \epsilon^\text{off}_R &= |\text{tr}(\rho_R A)-\langle A\rangle^\text{diag}_R| \label{eq:off_diag_error} \\
    \langle A\rangle_R^\text{diag} &= \sum_E
    \braket{E|A|E}\braket{E|\rho_R|E}
    \label{eq:diag_ens}.
\end{align}
\end{subequations}
The ``diagonal error" $\epsilon^\text{diag}_R$ captures the difference between the expectation value of $A$ in the microcanonical ensemble and the diagonal ensemble \cite{rigol_thermalization_2008} associated with $\rho_R$. It depends only on diagonal energy-basis matrix elements of $A$ and $\rho_R$. The ``off-diagonal error" $\epsilon^\text{off}_R$ captures error due to the fact that $\rho_R$ is not diagonal in the energy basis. Plugging Eq.~\eqref{eq:diag_ens} into Eq.~\eqref{eq:off_diag_error} yields 
\begin{equation}
    \epsilon^\text{off}_R = \sum_{E \neq E'} \braket{E|A|E'} \braket{E'|\rho_R|E}\,,
    \label{eq:error_off_diag}
\end{equation}
which involves only off-diagonal matrix elements of $A$ and $\rho_R$.

We now consider the dependence of the error $\epsilon_R$ on the ensemble size $R$, window size $\delta$, and system size $N$, assuming the ETH matrix element ansatz~\eqref{eq:eth}. 

\subsubsection{Diagonal error $\epsilon^\text{diag}_R$}
We begin with the diagonal contribution  $\epsilon^\text{diag}_R$. Plugging in the ETH ansatz to the expression for the diagonal error gives
\begin{multline}\label{eq:explicit_diagonal_error}
    \epsilon^{\rm diag}_R = \bigg| \sum_E \braket{E|\rho_{\rm mc} - \rho_R|E} A(E) \\
    + \sum_E \braket{E|\rho_{\rm mc} - \rho_R|E} \mathcal{D}^{-1/2}(E) f(E,0) R_{EE} \bigg|.
\end{multline}
The second line is $O(\mathcal{D}^{-1/2}(\lambda))$ \footnote{Technically, the function $f(\bar{E},\omega)$ can have some $N$ dependence for $|\omega| < O(N^{-1})$ but is expected to be $N$ independent at larger $|\omega|$; see the supplementary materials of \cite{dymarsky_new_2019} and see \cite{dalessio_quantum_2016} where $|f(0,\omega)|$ is argued to increase as $N^{1/2}$ in the former regime. For $\bar{E}$ away from zero, less appears to be known about its $N$ dependence, but in any case we neglect it, assuming that here the exponentially large density of states will suppress any $N$ dependence of $f(E,0)$ in the thermodynamic limit.} since both density matrices have unit trace. The density of states is evaluated at $\lambda$ since this is a typical energy in $W$. Via the triangle inequality,
\begin{equation}
    \epsilon^{\rm diag}_R \leq \big| \text{tr}[(\rho_{\rm mc} - \rho_R) A(H)]\big| + O(\mathcal{D}^{-1/2}(\lambda))
\end{equation}
where we have made the first term more compact by writing $A(H) = \sum_E A(E) \ket{E}\bra{E}$. Now we expand the smooth ETH function $A(E)$ near the target energy $\lambda$. Repeated uses of the triangle inequality yields
\begin{multline}
    \big| \text{tr}[(\rho_{\rm mc} - \rho_R) A(H)]\big| \leq \big| A(\lambda) \text{tr}[\rho_{\rm mc} - \rho_R ]\big| \\
    + \big| (\mathrm dA/\mathrm dE)(\lambda)\text{tr}[(\rho_{\rm mc} - \rho_R) (H-\lambda)]\big| + \cdots
\end{multline}
where the ellipsis signifies higher order derivatives. Both density matrices have unit trace, so the first term vanishes. But then using the fact that the $k^{\rm th}$ derivative of $A(E)$ with respect to $E$ is proportional to $N^{-k}$, which follows from $A(E) = a(E/N)$ with $a(x)$ becoming $N$-independent in the thermodynamic limit, we see that
\begin{equation}\label{eq:more_accurate_estimate}
    \epsilon^\text{diag}_R \leq \frac{\chi_R}{N} + O(N^{-2}) +  O(\mathcal{D}^{-1/2}(\lambda)),
\end{equation}
where
\begin{equation}
\label{eq:chi_R}
    \chi_R = \bigg| a'\bigg(\frac{\lambda}{N}\bigg)\text{tr}[(\rho_R-\rho_\text{mc})(H-\lambda)] \bigg|
\end{equation}
and $a'(x) = \mathrm da/\mathrm dx$. If both $\rho_R$ and $\rho_{\rm mc}$ have support only on the microcanonical window $W$, then $\chi_R \leq 2\delta |a'(\lambda/N)|$ for any $R$, where we have used that $|\text{tr}[\rho_R - \rho_{\text{mc}}]|\leq 2$. If instead they have support on a larger energy interval which is still $O(\delta)$, then $\chi_R$ is still $O(\delta)$ and we can still make the rough estimate
\begin{equation}\label{eq:estimate}
    \epsilon^\text{diag}_R \leq O(\delta/N) + O(\mathcal{D}^{-1/2}(\lambda)).
\end{equation}
In a very large system, we would not concern ourselves with the difference between Eq.~\eqref{eq:estimate} and the more accurate Eq. ~\eqref{eq:more_accurate_estimate}. However, at the relatively small systems we consider in this work, we can expect that for large $R$, $\chi_R$ and therefore $\epsilon^{\rm diag}_R$ will be smaller than $2\delta |a'(\lambda/N)|$ if we compare the variational estimate $\langle A \rangle^{\text{diag}}_R$ to the expectation value of $A$ in a microcanonical ensemble that is similar to the diagonal variational ensemble. In Sec.~\ref{section:numerical} we numerically confirm this for a Gaussian microcanonical ensemble. It is interesting to note that any sub-extensive choice of $\delta$ will lead to vanishing diagonal error in the thermodynamic limit; this is simply a manifestation of statistical-mechanical ensemble equivalence from the perspective of ETH.

\subsubsection{Off-diagonal error $\epsilon^\text{off}_R$}
We now turn to the off-diagonal error $\epsilon^{\rm off}_R$. First we observe that if the states $\ket{\psi_r}$ have exactly zero energy weight outside the microcanonical window $W$, the off-diagonal error \eqref{eq:off_diag_error} can be expressed as $\epsilon^\text{off}_R = | \text{tr}\rho_R \tilde{A}|$, where
\begin{align}
\label{eq:A-tilde}
    \braket{E|\tilde A|E'} =
    \begin{cases}
    \braket{E|A|E'} \ &\text{if}\ E,E'\in W \ \text{and} \ E\neq E' \\
    \quad\ \  0 \quad &\text{otherwise}
    \end{cases}.
\end{align}
If the variational states have some non-zero weight outside $W$, we can expect that the error is still approximately expressible this way, by slightly expanding $W$. In Sec.~\ref{section:numerical}, we suitably modify the expression  $\epsilon^\text{off}_R = |\text{tr}\rho_R \tilde{A}|$ in this case.
Since we are interested in understanding how averaging over an $R$-state ensemble can reduce this error, we write the average explicitly as
\begin{equation}\label{eq:error_dephase}
    \epsilon^\text{off}_R = \bigg\lvert \frac{1}{R}\sum_r \braket{\psi_r|\tilde{A}|\psi_r} \bigg\rvert.
\end{equation}
Following Refs.~\cite{dymarsky_new_2019,dymarsky_bound_2022}, let $x_r = \braket{\psi_r|\tilde{A}|\psi_r}$. For each $r\in\{1,2,..,R\}$ we have that
\begin{equation}\label{extreme_eigenvalues}
    \lambda_\text{min}(\tilde{A}) \leq x_r \leq \lambda_\text{max}(\tilde{A})
\end{equation}
where the limits are the minimum and maximum eigenvalues of the (purely off-diagonal) operator $\tilde{A}$. We find numerically in Sec.~\ref{section:truncate} Fig.~\ref{fig:sing_val} that the maximum/minimum eigenvalues of various operators are always above/below zero, respectively, which is expected since $\tilde{A}$ is traceless by construction. If the ensemble of microcanonical superposition states is ``random enough" we can expect that for each $r$, $x_r$ fluctuates between these limits according to some distribution. Should the algorithm perform ideally, the states $\ket{\psi_r}$ would sample $\tilde{A}$ in an unbiased way; i.e.~in the limit of infinite samples, $\epsilon^\text{off}_R\rightarrow \text{tr}(\tilde
{A})/n=0.$ This would happen for example if $\ket{\psi_r}$ were drawn from the Haar measure on the microcanonical subspace.

However, let us allow for biased sampling by modelling $x_r$ as identical and independently distributed (IID) random variables with covariance $\mathbb{E}(x_r x_s)-\mathbb{E}(x_r)\mathbb{E}(x_s)=\sigma^2\delta_{rs}$ and mean $\mathbb{E}(x_r) = c$ which would ideally be zero. With $x_r$ modelled this way, a simple statistical measure of the size of the off-diagonal error would be its mean-square value which has the functional form
\begin{equation}
    \mathbb{E}[(\epsilon^{\rm off}_R)^2] = \frac{\sigma^2}{R} + c^2 \equiv y_{c,\sigma}(R).
\end{equation}
For large $R$, a good approximation to the actual variance $\sigma^2$ should be given by the finite-size estimate
\begin{equation}
    \sigma_R^2 = \frac{1}{R}\sum_r x_r^2 - \bigg(\frac{1}{R}\sum_r x_r\bigg)^2,
\end{equation}
which is bounded from above by the largest square singular value of $\tilde{A}$ which is in turn shown in Appendix~\ref{section:truncate} to scale down weakly with $N$. Under such assumptions, the off-diagonal error in the VME algorithm should thus behave typically as
\begin{equation}
    \epsilon_R^{\rm off} = |c| + O(R^{-1/2})
\end{equation}
where we have assumed that $\sigma \approx \sigma_R = O(1)$. In Sec.~\ref{section:diagonal_error} we demonstrate that $x_r$ are indeed well modelled by this phenomenological description with some small non-zero $|c|$ always present. This implies that the variational states indeed are not sampling the microcanonical Hilbert space perfectly uniformly, which is consistent with them also not being close to Haar-random states.

\subsubsection{Summary}
In summary, the basic idea behind our algorithm is as follows. To prepare an approximation to $\rho_\text{mc}$, instead of preparing an ensemble of one or more eigenstates $\ket{E}$, which each individually require exponential quantum resources to generate, we will prepare a polynomially large number $R$ of states $\ket{\psi_r}$ which each hopefully require only polynomial quantum resources to generate, such that $\rho_R\approx\rho_\text{mc}$ \emph{as measured by local observables}. 

The error in the approximation can be understood in terms of two pieces. The first is the diagonal error, which is ultimately about statistical-mechanical ensemble equivalence as it manifests for an isolated quantum system via the diagonal part of the ETH ansatz \eqref{eq:eth}. The leading contribution to this type of error thus scales as $O(\delta/N)$, and thus any sub-extensive window width $\delta$ will in principle work. The more prohibitive error is the off-diagonal error, which for a single variational state the ETH alone cannot guarantee to be small at the scale of $\delta$ and $N$ practically accessible to the VME algorithm discussed in Sec.~\ref{section:algorithm}. To remedy this, we propose to insert randomness by averaging over variational states which have been prepared by initializing the VQA with random product states. We have so far focused on a particular observable $A$ and discussed the error in the context of its matrix elements. The claim that $\rho_R\approx\rho_\text{mc}$ \emph{as measured by local observables} can be made more precise by introducing the trace distance between certain reduced density matrices, which we examine numerically in Sec.~\ref{section:numerical}.

\subsection{VME algorithm}
\label{section:algorithm}

\IncMargin{1em}
\begin{algorithm}\label{algorithm}
    \KwData{Hamiltonian $H$, target energy $\lambda$, tolerance $\delta$ and random product state $\ket{\psi^0_r}$}
    \KwResult{converged variational state $\ket{\psi_r}$}
    \BlankLine
    $p=1$\;
    $\bm{\theta}=\bm{0}$\;
    \While{$p < \infty$}{
        $\epsilon = 10 $\;
        \While{$\epsilon \geq 10^{-3}$}{
            \While{$||\bm{\nabla} \mathcal{C}||_{\infty} > \epsilon$}{
            prepare $\ket{\psi_r(\bm{\theta})}=U_p(\bm{\theta})\ket{\psi^0_r}$\;
            measure $\mathcal{C(\bm{\theta})}$ and $\{\partial_j \mathcal{C(\bm{\theta})}\}_j$ \;
            update $\bm{\theta}$ according to BFGS optimizer\;
            }
            \eIf{$\mathrm{Var}(H) \leq \delta^2$}{
                $p^* \gets p$\;
                $\bm{\theta^*} \gets \bm{\theta}$\;
                converged\;
                set $\ket{\psi_r} = U(\bm{\theta}^*)\ket{\psi^0_r}$\;
            }{
                $\epsilon \gets \epsilon/2$\;
            }
        }
        $p \gets p+1$\;
    }
    \caption{prepare  $\ket{\psi_r(\bm{\theta^*})}$}\label{algo:searchmin}
\end{algorithm}\DecMargin{1em}

We now describe the VME algorithm for preparing microcanonical superposition states $\ket{\psi_r}$ discussed in the previous section. At the beginning we fix a target energy $\lambda$, and microcanonical window size
\begin{align}
\label{eq:delta}
\delta = \frac{\Delta E}{N} N^{\alpha},    
\end{align}
where $\Delta E$ is the full energy bandwidth of the Hamiltonian $H$. In the mixed-field Ising model we later consider, we find $\frac{\Delta E}{N}\approx 3$ independent of $N$. We focus our numerical studies mainly on the case $\alpha=-1/2$. We initialize the QPU (see Fig.~\ref{fig:high_level}) in a random product state
\begin{equation}
    \ket{\psi^0_r} = \ket{\varphi^1_r}\ket{\varphi^2_r}\cdots\ket{\varphi^N_r},
\end{equation}
with $\ket{\varphi^j_r} = \cos(\varphi^j_r)\ket{0} + \sin(\varphi^j_r)\ket{1}$ and $\varphi^j_r$ drawn from the uniform distribution on $[0,\pi)$, which we can expect to have extensive energy variance \cite{rigol_thermalization_2008,banuls_entanglement_2020}. We then minimize the ``folded-spectrum" cost function~\cite{peruzzo_variational_2014}
\begin{equation}\label{cost}
    \mathcal{C}(\bm{\theta}) = \braket{\psi (\bm{\theta})|(H-\lambda)^2|\psi(\bm{\theta})},
\end{equation}
until $\text{Var}(H)=\braket{(H-\braket{H})^2} \leq \delta^2$, obtaining the converged variational state $\ket{\psi_r}$. Note that $\mathcal C(\bm{\theta})$ penalizes both large energy variance and deviation of the average energy from the target energy, since
\begin{equation}
    \mathcal{C(\bm{\theta})} = \text{Var}(H) + \langle H-\lambda \rangle^2,
\end{equation}
but that the convergence criterion only concerns Var($H$). We find in practice that $\langle H-\lambda \rangle^2$ is comparatively small when $N$ is large, so it is also possible to think of $\mathcal{C}(\bm{\theta})\lesssim \delta^2$ as the convergence criterion.

The optimization is repeated $R$ times for different initial random product states to generate the variational ensemble. Notice that this cost function $\mathcal{C(\bm{\theta})}$ is zero if and only if $\ket{\psi(\bm{\theta})}=\ket{\lambda}$, the eigenstate with energy $\lambda$ \footnote{That is, assuming $\lambda$ is in the spectrum of $H$, or else the cost function is not zero but minimized for the nearest eigenstate, so that the cost function will still be exponentially small in the state $\ket{\lambda}$}. Unlike previous explorations \cite{zhang_adaptive_2021,shen_quantum_2017,mcclean_theory_2016} with this cost function, we do not seek local or global minima, since we minimize the cost function only until $\text{Var}(H) \leq \delta^2$, which is not a constraint on the gradient, but on the value of the cost function itself. Furthermore, the unique global minimum is a state completely different from the one we target. 

\begin{figure}
    \label{fig:ansatz}
    \includegraphics{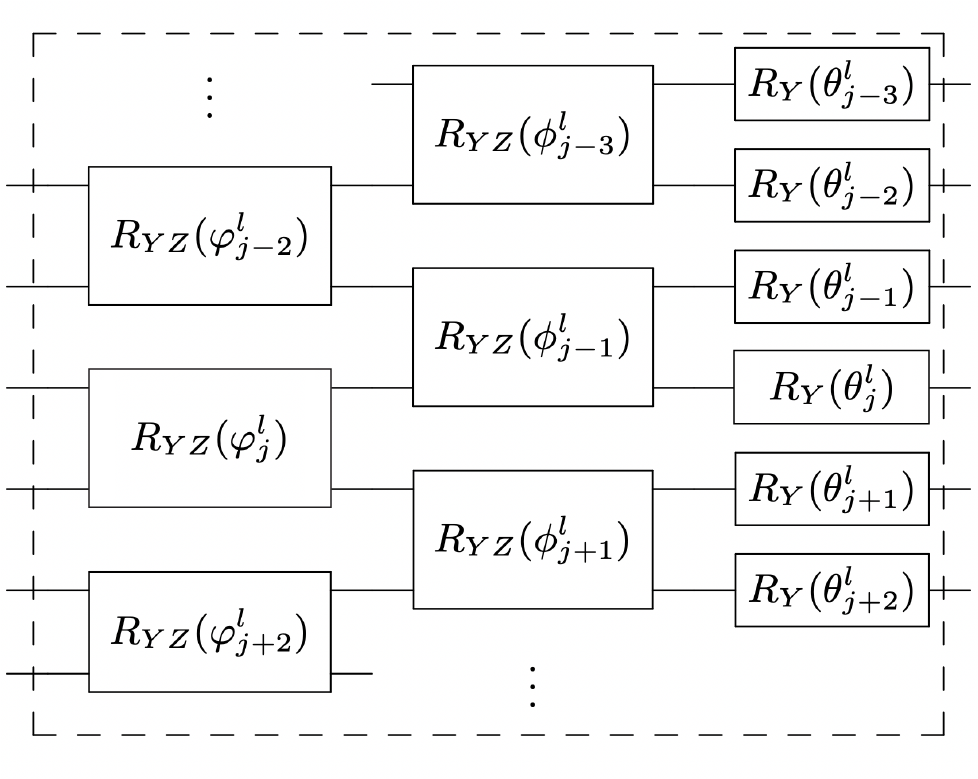}
    \setlength{\abovecaptionskip}{-5pt}
    \caption{Layer $l$ of the ansatz circuit, Eq.~\eqref{eq:single_layer}, near qubit $j$.}
\end{figure}

For simplicity we restrict our variational states to be real in the computational basis (CB). Because we are interested in the minimal circuit depth needed to prepare the variational states, we employ a periodic structure ansatz (PSA) \cite{larocca_diagnosing_2022} circuit for which the number of ``layers" will be adaptively chosen by the algorithm. The PSA with $p$ layers is defined as
\begin{equation}\label{eq:psa}
    U_p(\bm{\theta}) = \prod_{l=1}^p V(\bm{\theta}_l),
\end{equation}
where each layer is the unitary (see Fig. \ref{fig:ansatz})
\begin{equation}\label{eq:single_layer}
    V(\bm{\theta}_l) =
    \prod_{j=1}^N e^{i\theta_j^l Y_j}
    \prod_{\substack{j=1\\ \text{even}}}^N e^{i\phi_j^l Y_j Z_{j+1}}
    \prod_{\substack{j=1\\ \text{odd}}}^N e^{i\varphi_j^l Y_j Z_{j+1}}
\end{equation}
and $\bm{\theta}$ stands for the $2Np$ real parameters $\{\theta_j^l,\phi_j^l,\varphi_j^l\}_{jl}$ and $Y_j, Z_j$ are Pauli operators acting on qubit $j$. The ``brickwall" form of this ansatz breaks down for odd $N$, so in this case we add an additional gate to entangle the ends of the chain. That is, for odd $N$, we make the replacement
\begin{equation}
    \prod_{\substack{j=1\\ \text{even}}}^N e^{i\phi_j^l Y_j Z_{j+1}} \rightarrow 
     e^{i\phi_1^l Y_1 Z_N}
    \prod_{\substack{j=1\\ \text{even}}}^N e^{i\phi_j^l Y_j Z_{j+1}}.
\end{equation}

\begin{figure*}\label{fig:diagonal_error}
    \includegraphics{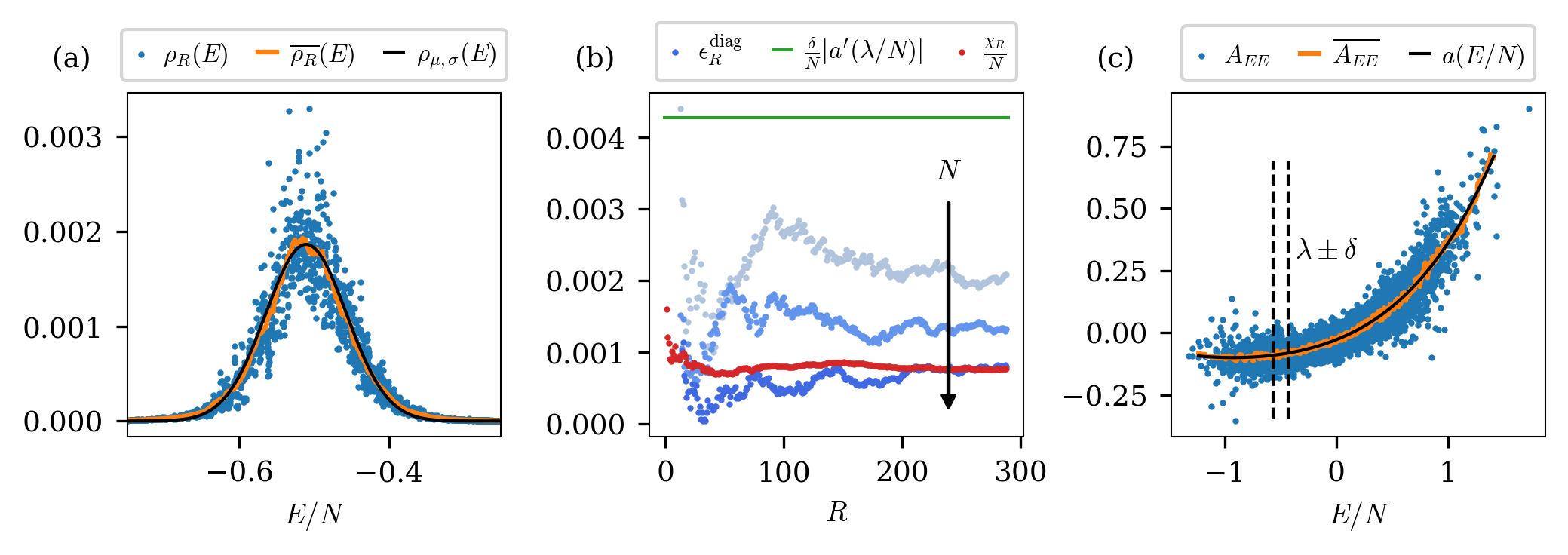}
    \setlength{\abovecaptionskip}{-5pt}
    \caption{Analysis of diagonal error in the VME, with operator $A=Z_{\floor{N/2}}$ taken as an example. In panel (a), the variational ensemble diagonal matrix elements $\rho_R(E)=\braket{E|\rho_R|E}$ (blue) for $N=13$, $\lambda/N=-0.5$, $\alpha = -1/2$ [see Eq.~\eqref{eq:delta}], and $R=288$ states in the ensemble. In black, the best Gaussian fit curve $\rho_{\mu,\sigma}(E)$ [see Eq.~\eqref{eq:broad_ens}], and in orange a coarse-grained version of the blue scatter points for comparison. Panel (b) shows the diagonal error [see Eq.~\eqref{eq:diag_error}] with $\rho_{\text{mc}} = \rho_{\lambda, \delta}$ versus $R\geq12$ for $N=11,12,13$ where increasing $N$ corresponds to darker blue data points. For comparison, we include in green the ensemble-independent estimate $\delta|A'(\lambda)|=\delta |a'(\lambda/N)|/N$ at $N=13$, as well as the more accurate estimate $\chi_R$ [Eq.~\eqref{eq:chi_R}] at $N=13$. Panel (c) includes the diagonal matrix elements $A_{EE} = \braket{E|A|E}$ in blue, their coarse-graining in orange, and the best fourth-order polynomial fit in black which defines $a(E/N)$. This smooth function $a$ is used to compute $\chi_R$. Vertical dashed lines show the scale of the microcanonical window as compared to the whole spectrum.}
\end{figure*}

The operators appearing in the single layer unitary $V(\bm{\theta}_l)$ are chosen based on the findings of Ref.~\cite{tang_qubit-adapt-vqe_2021}. There it is argued that the pool of $2N-2$ operators $\mathcal{P} = \{iY_jZ_{j+1}\}_{j=1}^{N-1} \cup \{iY_j\}_{j=1}^{N-1}$ is ``complete" in the sense that for any state $\ket{\psi}$, the set of states $\{A_k\ket{\psi}\}_k$ form a complete basis, where $A_k$ are nested commutators of operators in $\mathcal{P}$, i.e. elements of the dynamical Lie algebra \cite{dalessandro_introduction_2007} of $\mathcal{P}$. We have added the extra gates $Y_N$ and (for odd $N$) $Y_1Z_N$ to the pool, but clearly $\mathcal{P}\cup \{Y_1 Y_N,Y_1\}$ is still complete in the above sense.


Since our convergence criterion is based on the value of the cost function and the native convergence criterion of a gradient based optimizer is based on the size of the gradient, we ``wrap'' the optimizer in a simple loop (see Algorithm \ref{algorithm}) where we repeatedly interrupt the optimizer to check if the convergence criterion is satisfied, which we accomplish by having it only minimize until the gradient norm falls below a relatively large value $\epsilon$ which starts at $10^1$ and can only be decreased down to $10^{-3}$. If the algorithm then cannot achieve convergence using a $p$-layer ansatz by decreasing $\epsilon$ to $10^{-3}$, it adds another layer $p \rightarrow p+1$ and repeats the procedure. 

For the classical optimizer we employ the Broyden-Fletcher-Goldfarb-Shanno (BFGS) optimizer which is gradient-based. We therefore take advantage of the ``parameter shift rule" \cite{schuld_evaluating_2019} for computing analytic gradients of the cost function. If the generators are Pauli strings (hence squaring to $I$), then  
\begin{equation}
    \frac{\partial}{\partial\theta_j} \mathcal{C}(\bm{\theta}) = \mathcal{C}(\bm{\theta}+\frac{\pi}{4} \bm{e}_j) - 
    \mathcal{C}(\bm{\theta}-\frac{\pi}{4} \bm{e}_j)
\end{equation}
with $\bm{e}_j$ a unit vector in the $j^{\rm th}$ direction. Thus, during the optimization both the cost function and its derivatives can be measured using the QPU.


\section{Numerical Results}\label{section:numerical}

We test the VME algorithm on the 1D Mixed Field Ising Model (MFIM) with Hamiltonian
\begin{equation}\label{eq:mfim}
    H = \sum_{j=1}^N \left( J Z_j Z_{j+1} + h_{x,j} X_j  + h_z Z_j \right)
\end{equation}
and periodic boundary conditions so that site $N+1$ refers to site 1. To ensure there are no accidental degeneracies we consider weakly nonuniform transverse fields $h_{x,j} = h_x + r_j$, where $r_j\in[-0.01,0.01]$ are drawn randomly from the uniform distribution. We use a single fixed configuration of the transverse fields for our numerics. We fix parameters $J = 1$, $h_z=0.5$ and $h_x = -1.05$ such that the system is strongly nonintegrable \cite{banuls_strong_2011}. In this section we discuss the performance of the VME with respect to this particular model.

\subsection{Diagonal variational ensemble}\label{section:diagonal_error}
Here we characterize the nature of the converged variational states in the energy basis. To do so, we first define a ``broadened'' microcanonical ensemble as was done in Ref.~\cite{schrodi_density_2017}. This ensemble is of the form
\begin{equation}\label{eq:broad_ens}
    \rho_{\lambda,\delta} = \mathcal{D}^{-1}_\delta (\lambda)G_\delta(H-\lambda)
\end{equation}
where $G_\delta(x) = (2\pi\delta^2)^{-1/2}e^{-x^2/2\delta^2}$ is a normalized Gaussian function, and $\mathcal{D}_\delta (\lambda)=\text{tr } G_\delta(H-\lambda)$ is the ``broadened" density of states evaluated at energy $\lambda$. Here, $\delta$ corresponds to the convergence criterion for the VME algorithm, i.e.~Eq.~\eqref{eq:delta} with $\alpha=-1/2$. We will from now on treat $\mathcal{D}_\delta(\lambda)$ as a good approximation to the density of states in the thermodynamic limit (see Appendix~\ref{section:dos} for further justification).

We claim that the variational algorithm~\ref{algo:searchmin} generates diagonal energy-basis matrix elements $\rho_R(E) = \braket{E|\rho_R|E}$ approximating a broadened microcanonical ensemble. Fig.~\ref{fig:diagonal_error}(a) shows the variational ensemble diagonal energy-basis matrix elements to which we fit the curve $\rho_{\mu,\sigma}(E) = \mathcal{D}^{-1} (\mu)G_\sigma(E-\mu)$ 
with fitting parameters $\mu$ and $\sigma$ (shown in solid black). Up to fluctuations from eigenstate to eigenstate, we can see that the variational diagonal ensemble is well described by the Gaussian best-fit. More precisely, the coarse grained version, $\overline{\rho_R}(E)$, of the variational ensemble diagonal elements–where the fluctuations are eliminated–agrees quite well with the Gaussian best-fit. The coarse-grained curve is computed as follows: for each $E$ in the spectrum of $H$, $\overline{\rho_R}(E)$ is defined as the average of $\braket{E'|\rho_R|E'}$ over the $K$ eigenenergies $E'$ nearest to $E$. We set the ``resolution" $K=64$, except for $E$ near the edges of the spectrum where $1\leq K<64$. At the largest system size of $N=13$ and for an $R=288$-state ensemble, we list the best fit parameters $\mu$ and $\sigma$ in Table \ref{tab:converged_ensembles}.

\begin{table}[b]
\centering
\begin{tabular}{*{3}c}
\toprule
$\lambda/N$ & $\mu/N$ & $\sigma/\delta$ \\
\midrule
$-0.750$ &  $-0.761$ & $0.858$  \\
$-0.500$ & $-0.511$ & $0.831$ \\
$-0.250$ & $-0.253$ & $0.821$ \\
$0.000$  & $0.001$ & $0.832$\\
\bottomrule
\end{tabular}
\caption{The $N=13$ broadened microcanonical best fit parameters $(\mu,\sigma)$ at various target energy densities $\lambda/N$.}
\label{tab:converged_ensembles}
\end{table}


Note that away from $\lambda=0$, the variational ensembles converge with $\mu$ slightly different than the target $\lambda$. This is because the variational ensemble is generated by minimizing the first two moments of the operator $(H-\lambda)$, but the density of states determined by the underlying model is non-uniform. In fact if we assume the Gaussian density of states discussed in Appendix~\ref{section:dos}, we have $\text{tr}(\rho_{\mu,\sigma}H) \approx \mu -O(\sigma^2\frac{\mu}{N})$, where we have assumed that $\sigma$ decreases with $N$ so that higher order terms can be neglected in the thermodynamic limit. We can see that when $\mu<0$, the $\rho_{\mu,\sigma}$ ensemble actually has average energy larger than $\mu$. In Appendix~\ref{section:var_ens} we confirm this statement more quantitatively. As a consequence of this analysis, we conclude that the deviations in $\mu$ from $\lambda$ are a finite-size effect due to the non-constant density of states, and not due to the fluctuations around the average curve $\overline{\rho_R}(E)$. 

In the last row of Table~\ref{tab:converged_ensembles}, we see that all the $\sigma$ are at least about $15\%$ smaller than $\delta$. This is due to a simplification we have made in the preceding analysis. Note the form of $\overline{\rho_R}(E)$ in Fig.~\ref{eq:diag_error}(a) at the edges of the window; the orange curve has more weight away from the window than the Gaussian best-fit (in black). A more accurate characterization of the ensemble might be, for example, a sum of two Gaussian curves. We nonetheless opt to consider the ensemble as roughly a single Gaussian peak for simplicity. We discuss the quantitative consequences of this simplification in Appendix~\ref{section:broadened} and explain why the $\sigma/\delta$ shown in Table~\ref{tab:converged_ensembles} are not closer to unity. 

Up to fluctuations around the coarse-grained behavior $\overline{\rho_R}(E)$ and the slight oversimplification of the single peak Gaussian fit, we have thus established the form of the diagonal part of the variational ensemble. In the following sections we will study the error in the VME estimate of various observables. Clearly, we will need to choose an appropriate microcanonical ensemble to compare to. This ensemble is arguably $\rho_{\mu,\sigma}$ because it closely approximates the (coarse-grained) diagonal ensemble $\overline{\rho_R}(E)$. One could also imagine comparing to the diagonal ensemble itself and focusing solely on the off-diagonal error as was done in \cite{cakan_approximating_2021}. However, imagining the VME as a practical algorithm for computing broadened microcanonical ensemble averages, one could not know ahead of time what $\mu$ and $\sigma$ were. Furthermore we have shown that the distinction between $\lambda/N$ and $\mu/N$ is a finite-size effect that is already small at $N=13$. We will henceforth compare our numerical results to the ensemble $\rho_{\lambda,\delta}$, where $\lambda$ and $\delta$ are precisely the parameters that were initially chosen before running the algorithm, i.e. we set $\rho_{\rm mc} = \rho_{\lambda,\delta}$ in Eq.~\eqref{eq:diag_error}.

\subsection{Diagonal error}
\label{subsec:diagonal_error}
We now briefly discuss the diagonal error with respect to the broadened microcanonical estimates. Fig.~\ref{fig:diagonal_error}(b) shows, for $N=11,12,13$, the diagonal error versus $R\geq12$ for the operator $A=Z_{\floor{N/2}}$ at $\lambda/N=-0.5$. For comparison we plot a simple estimate of the error $(\delta/N)|a'(
\lambda/N)|$ at $N=13$, as well as the more accurate estimate $\chi_R$ defined via Eq. \eqref{eq:more_accurate_estimate} which we calculate numerically using the $N=13$ variational ensemble. We calculate $a'(\lambda/N)$ using a fourth-order polynomial fit to the (coarse-grained) graph of $\bra{E} A \ket{E}$ versus $E/N$, shown as the solid black line in panel (c) of Fig.~\ref{fig:diagonal_error}. The coarse-grained curve $\overline{A_{EE}}$ is computed in the same way as was $\overline{\rho_R}(E)$ in Sec.~\ref{section:diagonal_error}, but here we use a resolution of $K=32$.

For this particular operator and energy density, it is clear that the diagonal error decays with $N$ and is an order-one fraction of the rough estimate $O(\delta/N)$. Furthermore we can see that for large $R$ its behavior is well captured by the expected estimate $\chi_R$.

In Appendix~\ref{section:diagonal_data} we present further numerical results for the four local operators $Z,ZZ,X,XX$ acting on the central one or two sites of the chain at the energy densities $\lambda/N=-0.75,-0.5,-0.25,0.0$. At $\lambda/N=-0.5$, all operators have the property that $\epsilon^\text{diag}_R$ decays with $N$ for large $R$, and the $N=13$ values are consistent with the estimate $\chi_R/N$. At other energy densities the scaling with $N$ is not always so well established, but the error for large $R$ is always smaller than $(\delta/N)|a'(\lambda/N)|$. The value of $\chi_R$ also appears to generally be on the correct scale of $\epsilon^\text{diag}_R$ except for the operator $XX$, for which $\chi_R/N$ undershoots the value of the diagonal error for the higher energy densities. These various deviations are likely due to ETH not yet strongly setting in at such small system sizes. In particular the $N=13$ value of $\mathcal{D}^{-1/2}(\lambda)$ is never smaller than $0.04$ for the energy densities we consider, so that the ETH fluctuations, i.e. the third term in equation \eqref{eq:more_accurate_estimate}, could be comparable to $\delta/N\approx 0.06$ at $N=13$ depending on the relative size of $A(E)$ and the ETH function $f(E,0)$. 

\begin{figure*}
\label{fig:off_diagonal_error}
    \centering
    \includegraphics{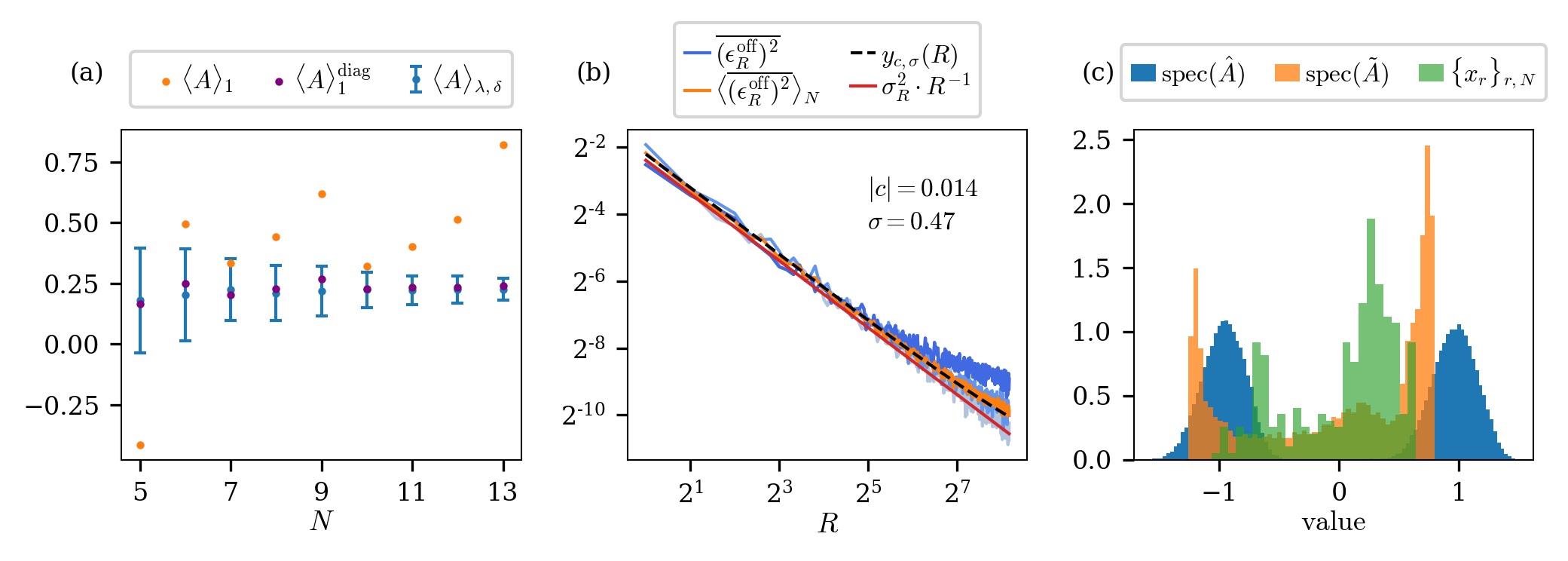}
    \setlength{\abovecaptionskip}{-5pt}
    \caption{Off-diagonal error in the VME algorithm, with $A=X_{\floor{N/2}}$ at $\lambda/N=-0.5$ taken as an example. In panel (a), the broadened microcanonical ensemble expectation value $\langle A\rangle_{\lambda, \delta} = \text{tr}(\rho_{\lambda, \delta} A)$ and error bars indicating the associated broadened microcanonical standard deviations (obtained from ED) is plotted in blue as a function of $N$. We compare this to the estimate in a single variational state, $\langle A\rangle_1=\text{tr}(\rho_1 A)$, along with the corresponding \textit{diagonal ensemble} estimation $\langle A\rangle^{\text{diag}}_1$, i.e.~Eq.~\eqref{eq:diag_ens} with $R=1$. Note that $\langle A\rangle^{\text{diag}}_1$ can only be computed with ED, where off-diagonal contributions can be discarded by hand. In panel (b) we show the finite-sample mean-square off-diagonal error $\overline{(\epsilon_R^\text{off})^2}$ with increasing $R$ for $N=11,12,13$, where increasing $N$ corresponds to a darker blue curve as in Fig.~\ref{fig:diagonal_error}. The orange curve, $\langle \overline{(\epsilon^{\rm off}_R)^2}\rangle_N$, is the $N$-averaged value of $\overline{(\epsilon_R^\text{off})^2}$ as discussed in the text. The dashed black line is a two-parameter best-fit to $\langle \overline{(\epsilon^{\rm off}_R)^2}\rangle_N$, and for comparison we plot the ``theoretical'' error $\sigma^2_R/R$ in red. Panel (c) shows probability density functions of the $N=13$ eigenvalues $\mathrm{spec}(\hat{A})$ and $\mathrm{spec}(\tilde{A})$ with $\hat{A}$ and $\tilde{A}$ as defined in the text. Panel (c) also shows a probability density function of the collection $\{x_r\}_{r,N}$ where $N$ runs over a few system sizes as discussed in the text.}
\end{figure*}

In any case, the diagonal error is always quite small across all operators and energy densities, when compared for example against the scale of the microcanonical fluctuations themselves. For example, see Fig.~\ref{fig:off_diagonal_error}(a) where in purple we can see that even for a \textit{single} variational state, the diagonal ensemble estimate is highly accurate. We observe this across every considered operator and energy density, as shown in Appendix~\ref{section:off_diagonal_data}.

\subsection{Off-diagonal error}
We now turn to discuss the numerical details of the off-diagonal error and how ensemble averaging reduces it. First, Fig.~~\ref{fig:off_diagonal_error}(a) illustrates the main problem with using a single variational state with a large $\delta$. We take as an example the operator $A=X_{\floor{N/2}}$ at the energy density $\lambda/N=-0.5$ and compare a single variational state estimate $\langle A \rangle_1 = \braket{\psi_1| A|\psi_1}$ to the smooth microcanonical average $\langle A \rangle_{\lambda,\delta} = \text{tr}(\rho_{\lambda, \delta} A)$. The estimate is poor even for the largest system size. In Fig.~\ref{fig:off_diagonal_error}(b) we examine how this error is reduced by averaging. It is clear that averaging always reduces the off-diagonal error–however since we anticipate it to behave as a random variable, we need to measure a statistical quantity to make the analysis precise. Given the finite dataset $\{x_r\}_{r=1}^{288}$ of samples, we calculate a finite size estimate of the mean-square error in whatever underlying distribution $x_r$ are sampled from as a function of $R$ as follows. For each fixed $R \in \mathcal{R} = \{1,2,\dots,288\}$ we calculate 
\begin{equation}\label{eq:scrambling}
    \overline{(\epsilon^{\rm off}_R)^2} = \frac{1}{S}\sum_{k=1}^S \bigg|\frac{1}{R} \sum_{r \in P_k(\mathcal{R})\vert_R} x_r \bigg|^2
\end{equation}
where $P_k(\mathcal{R})\vert_R$ is the first $R$ elements of a random permutation of the ordered index set $\mathcal{R}$. Choosing $S=100$, we plot $\overline{(\epsilon^{\rm off}_R)^2}$ versus $R$ for $N=11,12,13$ on a log-log scale in anticipation of observing an $R^{-1}$ scaling of the off-diagonal error. Across various operators and energy densities (shown in Appendix~\ref{section:off_diagonal_data}), we observe an initial $R^{-1}$ scaling with prefactor approximately independent of $N$. However, the large $R$ value can either be larger or smaller depending on $N$ in a non-systematic way. To remove this dependence we focus our analysis on a system-size averaged version
\begin{equation}
    \langle \overline{(\epsilon^{\rm off}_R)^2}\rangle_N = \frac{1}{5}\sum_{N=9}^{13} \overline{(\epsilon^{\rm off}_R)^2},
\end{equation}
which is shown in orange. The data is well described by a two parameter best fit function of the form $y_{c,\sigma}(R) = \sigma^2/R + c^2$ which corresponds to $x_r$ being effectively IID random variables as discussed in Sec.~\ref{section:var}. The values of $|c|$ and $\sigma$ are shown in panel (b). We can see that for the operator $X$ at $\lambda/N=-0.5$ shown here in the main text, the value of $|c|=0.014$ is quite small. In other cases, in particular for $XX$, it can be slightly larger: $|c|=0.071$. Data for all operators and energy densities are shown in Appendix~\ref{section:off_diagonal_data}. We also find that the best-fit parameter $\sigma$ is on the same scale as the finite-sample estimate $\sigma_R$ as demonstrated by the collapse of the red curve $\sigma^2_R/R$ and the best-fit line in the small $R$ regime. Here $\sigma_R$ is calculated for $\tilde{A}$ having non-zero energy-basis matrix elements only on an energy window of half-width $3\delta$  [see Eq.~\eqref{eq:A-tilde}]. This approximation of computing $\sigma_R$ based only on the matrix elements of $\rho_R$ and $A$ in energy eigenstates near $\lambda$ is numerically justified in Appendix~\ref{section:truncate}. These results imply that for small $R$, the statistics of the off-diagonal error are controlled by $\tilde{A}$.

Ref.~\cite{richter_eigenstate_2020} inferred correlations between energy-basis matrix elements of local operators $A$ by the form of the eigenvalue statistics of certain sub-matrices of $A$. To help understand the nature of the off-diagonal error, in Fig.~\ref{fig:off_diagonal_error}(c) we also examine the eigenvalue distribution of the operator $\tilde{A}$ defined on an energy window of half-width $3\delta$ and centered on energy $\lambda$, i.e. as in Eq.~\eqref{eq:A-tilde} but with $W$ slightly expanded to accommodate the tails of the roughly Gaussian variational states. See Appendix~\ref{section:truncate} for a graphical representation of how this energy window is defined. For comparison we also show in Fig.~\ref{fig:off_diagonal_error}(c) the eigenvalue distribution of the operator $\hat{A}$, which is simply the operator $A$ with its energy-basis diagonal elements deleted. There are a number of interesting qualitative properties displayed by these eigenvalue distributions that are relevant to the off-diagonal error in the VME.

Firstly, we observe that the eigenvalue distribution of $\tilde{A}$ does not appear to qualitatively change shape as $N$ is varied, except for a slight reduction in the total width for increasing $N$, as demonstrated in Appendix~\ref{section:truncate}. Since it is $\tilde{A}$ which roughly determines the off-diagonal error, the qualitative lack of $N$ dependence agrees with the fact that the off-diagonal error does not depend on $N$ in a systematic way at the system sizes we examine.

Interestingly, we observe a correlation between the single-state variational estimates in Fig.~\ref{fig:off_diagonal_error}(a) as $N$ is varied and the eigenvalue distribution of $\tilde{A}$. When $\langle A \rangle_1$ over/under-estimates the microcanonical value across many system sizes, the eigenvalue distribution is biased to the right/left of zero. For further evidence that this correlation is not an artifact of this energy density or the choice of operator, see Appendix~\ref{section:off_diagonal_data}. To demonstrate this further, we there also plot a histogram of the off-diagonal error present in many individual variational state samples, including samples across a window of system sizes: specifically we show a normalized histogram of the values in the set
\begin{equation}
     \{x_r\}_{r,N} = \bigcup_{N = 9}^{13} \{x^N_r\}_{r=1}^{288}
\end{equation}
where $x^N_r$ is the off-diagonal error in variational state $r$ when the system size is $N$. Using the statistics across multiple system sizes is justified here since the off-diagonal error varies erratically with the system size. On a qualitative level, this latter histogram confirms that the variational states sample $\mathrm{spec}(\tilde{A})$ uniformly enough that a bias in $\mathrm{spec}(\tilde{A})$ on the left or right of zero is reflected in the statistics of $x_r$. This fact is not too surprising since roughly speaking, $\tilde{A}$ determines the off-diagonal error via
\begin{equation}
    x_r \approx \sum_{\tilde{a}} \tilde{a} |\! \braket{ \tilde{a}|\psi_r}\! |^2
\end{equation}
with $\tilde{a}$ and $\ket{\tilde{a}}$ the eigenvalues and eigenvectors of $\tilde{A}$. However, it is not obvious that $|\braket{\tilde{a}|\psi_r}|^2$ is a uniform distribution and furthermore, this approximation should only be understood statistically since in actuality, it is the properties of $\hat{A}$ which precisely determine the error:
\begin{equation}
    x_r = \sum_{\hat{a}} \hat{a} |\! \braket{\hat{a}|\psi_r}\! |^2
\end{equation}
and the truncation of $\hat{A}$ to $\tilde{A}$ is only shown in Appendix~\ref{section:truncate} to rigorously hold for the quantity $\sigma_R$ as opposed to individual realizations $x_r$.

At all energy densities, the eigenvalue distribution of $\widetilde{XX}$ is much flatter than the other three considered operators, and the histogram of $x_r$ is also qualitatively different for $XX$, two observations which could be related to the fact that the off-diagonal error for $A=XX$ generally saturates sooner and to a larger value than the other operators do. However, we leave an identification of the precise underlying mechanism to future work.

We conclude the discussion of the off-diagonal error with some observations about the eigenvalue statistics of the full-spectrum off-diagonal operator $\hat{A}$. Even though the variational states in principle have support on the entire energy spectrum, we can see that it is the statistics of $\tilde{A}$ and not of $\hat{A}$ that are correlated with the off-diagonal error, further justifying the truncation to a local energy window. Interestingly, we can see that $\hat{A}$ still has a similar spectral form to that of a Pauli string with eigenvalues $\pm1$, but the otherwise highly degenerate peaks have been smeared out by removing the diagonal energy-basis elements. In Appendix~\ref{section:off_diagonal_data} we show the eigenvalue distributions of $\hat{A}$ for other $A$, and note that $XX$ looks the most similar to that of a Pauli string, i.e. its peaks have been broadened the least. When the window is reduced to the scale $\delta$, the distribution becomes less similar to that of a Pauli operator, and we can expect that as $\delta \rightarrow 0$, the spectrum approaches that of a random matrix, i.e. the semi-circle law \cite{wigner_distribution_1958,dymarsky_bound_2022,richter_eigenstate_2020}. The fact that the eigenvalue distribution is so far from a semi-circle law on the scale $\delta=O(N^{-1/2})$ provides further confirmation that $\braket{E|A|E'}$ are not effectively independently distributed and thus we cannot rely on randomness of the matrix elements alone to make the off-diagonal error small.

\subsection{Explicit microcanonical estimates and trace distance}

\begin{figure}\label{fig:observables}
    \centering
    \setlength{\abovecaptionskip}{-5pt}
    \setlength{\belowcaptionskip}{-5pt}
    \includegraphics{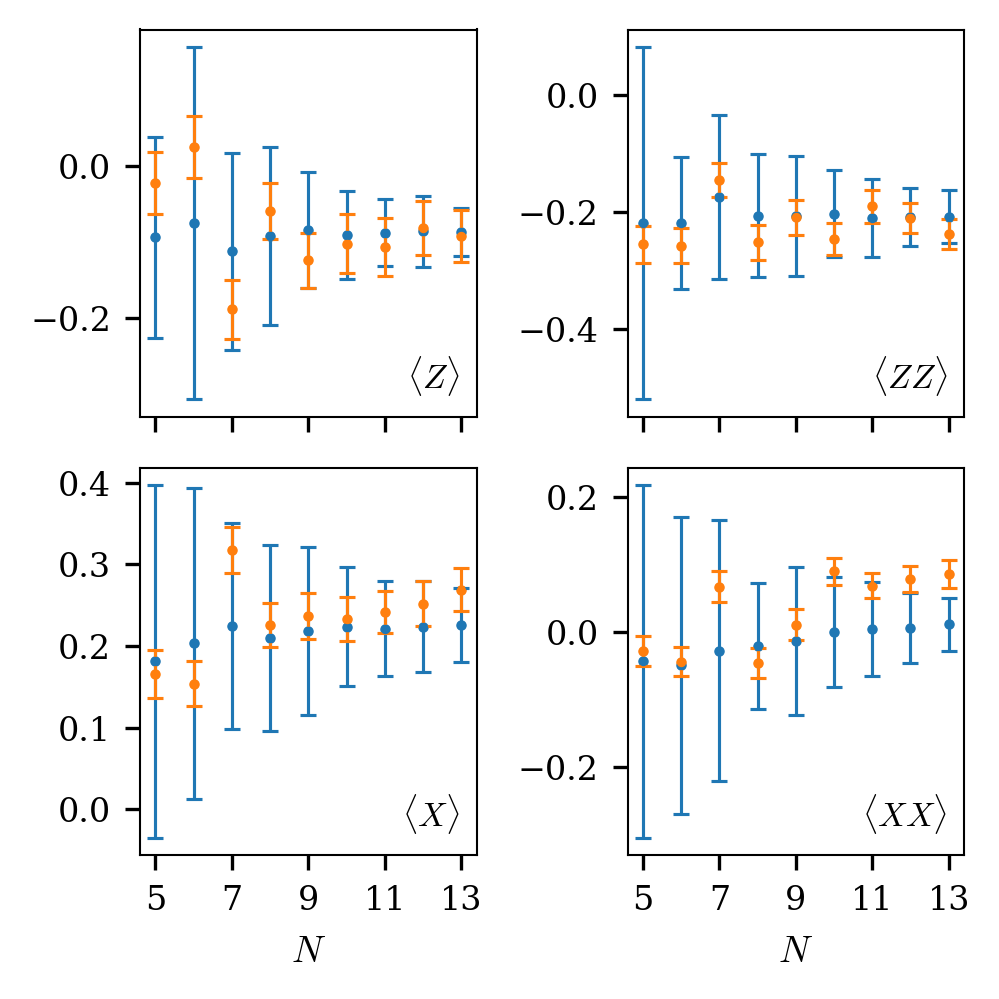}
    \caption{In blue, smooth microcanonical averages $\langle A \rangle_{\lambda,\delta} = \text{tr}(\rho_{\lambda, \delta} A)$ and their thermal fluctuations,  Eq.~\eqref{eq:mc_fluctuations}. In orange, the corresponding variational estimates for $\alpha =-1/2$ and $R=288$ with standard error. Averages and their error are plotted as a function of system size $N$ at a fixed energy density of $\lambda/N=-0.5$.}
\end{figure}

Having examined in some detail the scaling of the absolute diagonal and off-diagonal errors, we now take a step back and consider what the overall statistics of the variational estimates look like when compared to the microcanonical averages and their associated microcanonical fluctuations. For example, in Fig.~\ref{fig:observables} we show for various system sizes the variational estimate $\text{tr}(\rho_RA)$ for $R=288$ along with the standard error. This is compared against the broadened microcanonical average calculated from ED, with error bars indicating one microcanonical standard deviation $ \Delta A_{\lambda,\delta}$, i.e.,
\begin{equation}\label{eq:mc_fluctuations}
    (\Delta A_{\lambda,\delta})^2 = \sum_E \braket{E|\rho_{\lambda,\delta}|E} \braket{E|A|E}^2 - (\text{tr}\rho_{\lambda,\delta} A )^2
\end{equation}
which is another scale to which the error can be compared.

Running the VME two different times yields two different variational ensembles $\rho_R$ and $\rho'_R$. The orange error bars measure how much $\text{tr}(\rho_RA)$ and $\text{tr}(\rho'_RA)$ would differ when $R=288$. We find the energy densities $\lambda/N=-0.75$ and $\lambda/N=-0.5$ generally have more accurate estimates than $\lambda/N=-0.25$ and $\lambda/N=0$, see Appendix~\ref{section:micro_data} for further results. For a fixed $R$, the error certainly does not systematically decrease with $N$. In some cases, it appears to increase with $N$, for example $X$ at $\lambda/N=-0.5$, shown in Fig.~\ref{fig:observables}, or in some cases for $XX$ as discussed in Appendix~\ref{section:micro_data}. We do not consider these observations for a fixed $R$ at odds with the off-diagonal error analysis where we stated that over a large range of $R$ the error does not appear to systematically depend on $N$. We note that the operator $XX$ generally appears to deviate at the larger system sizes more than $Z,X,ZZ$ across various energy densities, which agrees with the previous observations that $XX$ behaves differently than the other operators. 


\begin{figure}
    \centering\label{fig:trace}
    \setlength{\abovecaptionskip}{-5pt}
    \setlength{\belowcaptionskip}{-5pt}
    \includegraphics{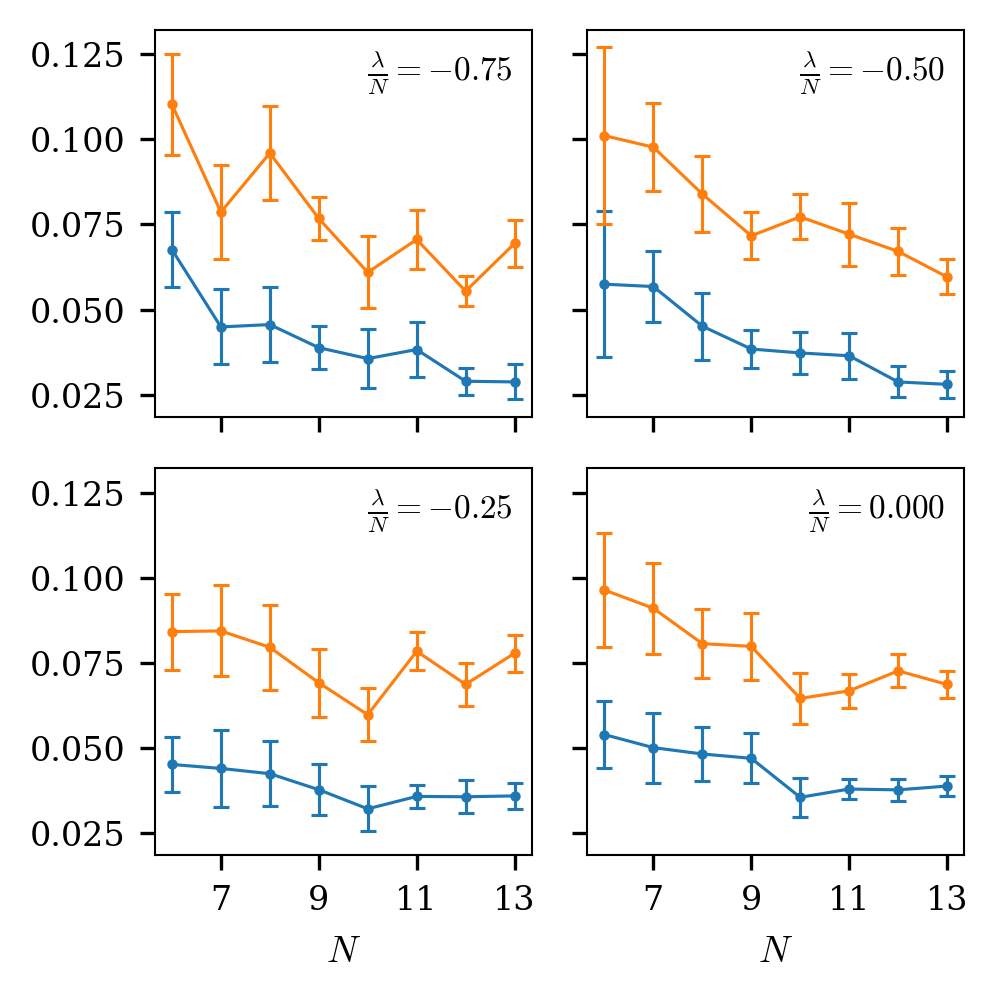}
    \caption{Spatially averaged trace distance between variational ensembles of size $R(N)=\lfloor1.5 N^2\rfloor$ and for $\alpha=-1/2$ and the corresponding broadened microcanonical ones. The blue curves are for $|S|=1$ and the orange ones for $|S|=2$. Error bars correspond to the standard deviation of this averaged quantity over $20$ different ensemble realizations.}
\end{figure}

\begin{figure*}\label{fig:resources}\includegraphics{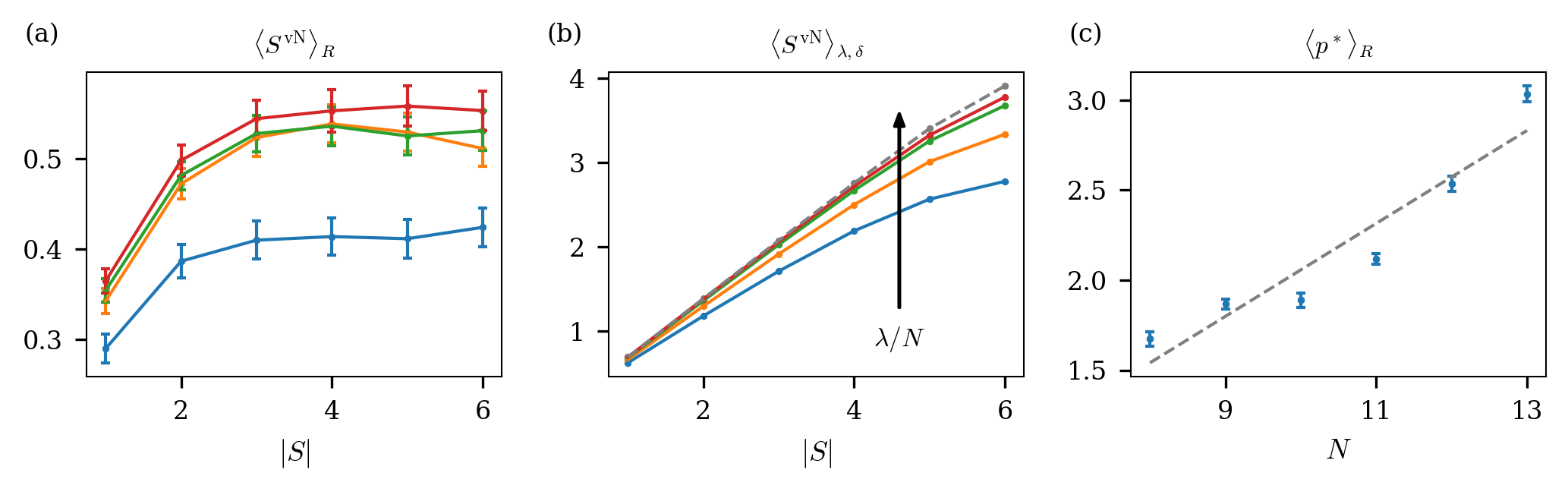}
    \setlength{\abovecaptionskip}{-5pt}
    \caption{Panels (a) and (b) show the entanglement entropy $S^{\rm vN}$ of a contiguous subregion $S$ for different energy densities, where the blue, orange, green, and red colored curves correspond to $\lambda/N = -0.75,-0.5,-0.25,0$, respectively. Panel (a) contains the ensemble average entanglement entropy of converged variational states at $N=13$ for an $R=144$ state variational ensemble with error bars indicating the standard error. For comparison, (b) shows the average entanglement entropy in the corresponding broadened microcanonical ensembles, see Eq.~\eqref{eq:smc}. The gray dashed curve is the Page entropy for a Haar-random state. Panel (c) graphs the number of layers in the PSA circuit for $\lambda/N=-0.5$, averaged over $R=144$ samples with error bars indicating the standard error. The dashed gray line is the best linear fit $p(N) = 0.26N-0.52$.}
\end{figure*}

To see if converged ensembles tend towards the microcanonical state with increasing $N$ in an observable-independent way, we check if, for a subsystem $S$ of the chain, the state $\rho_R^{S}=\text{tr}_{\bar{S}}(\rho_R)$ approaches the subsystem broadened microcanonical state $\rho_{\lambda,\delta}^{S}=\text{tr}_{\bar{S}}(\rho_{\lambda,\delta})$. As a distance measure we consider the trace distance, $T(\rho,\rho')=\frac{1}{2}||\rho-\rho'||_1$, which measures the distinguishability of $\rho$ and $\rho'$ in an operationally meaningful way \cite{wilde_quantum_2013}. More importantly for our purposes however, it also bounds the difference in the expected value of any observable $A$ as
\begin{equation}
    T(\rho,\rho')\geq \frac{|\text{tr}(\rho A) - \text{tr}(\rho' A)|}{2 \sigma_\text{max}(A)}\,.
\end{equation}
where $\sigma_\text{max}(A)$ is the maximum singular value of $A$. This can be derived by using von Neumann's trace inequality $|\text{tr}(XY)|\leq \sum_i \sigma_i(X)\sigma_i(Y)$. In particular for a Pauli string operator, we have $\sigma_\text{max}(A)=1$. We plot $T(\rho^S_R,\rho^S_{\lambda,\delta})$ versus $N$, averaged over all contiguous subsystems $S$ of fixed size $|S|$. We observe that the trace distance for a single site subsystem $|S| = 1$ is always smaller (by about a factor of $1/2$) than for a system of two nearest-neighbor sites $|S|=2$. In light of the previous analysis where the off-diagonal error did not appear to depend systematically on $N$, we choose $R$ to scale with $N$ as $R(N)=O(N^2)$, and further average over $20$ ensemble realizations in each case. Since we have only $288$ samples total, we compute the statistics of $20$ random (possibly overlapping) subsets of size $R$.

The reason one may want to increase $R$ with $N$ is that the trace distance satisfies the inequality
\begin{equation}\label{eq:trace_distance}
    T(\rho^S_R,\rho^S_{\lambda,\delta}) \leq \frac{1}{R} \sum_{r=1}^R  T(\rho^S_r,\rho^S_{\lambda,\delta}),
\end{equation}
with $\rho^S_r = \text{tr}_{\bar{S}}\ket{\psi_r}\bra{\psi_r}$, so that a large-$R$ variational ensemble can only do better than single pure states can on average. Since we are interested in understanding the behavior of the algorithm in the thermodynamic limit, and in light of our above results suggesting that the off-diagonal error does not appear to depend strongly on $N$, we consider the case of $R= O(N^2)$ so that we can observe a continual decrease of the trace distance with $N$ at the lower energy densities. It is possible however, that for a larger number of samples the spatially averaged trace distance will also eventually saturate to a finite value as it did for $A=(XX)_{\floor{N/2}}$ in particular.



The trace distance results in Fig.~\ref{fig:trace} are consistent with the estimates for particular local operators, for example when comparing Fig.~\ref{fig:observables} at $N=13$, we checked numerically that the deviation of the average estimate (corresponding to $R=288$) from the microcanonical one never exceeds twice the trace distance at the corresponding energy density. However, the $N=13$ value of the trace distances shown correspond to $R=253$, whereas the observables correspond to $R=288$. Furthermore, the trace distance is averaged over all contiguous subsystems of size $|S|$ so in principle this value no longer exactly upper bounds the observables as in Eq.~\eqref{eq:trace_distance}, which are obtained for a given site $j = \floor{N/2}$, but in this case the results are nonetheless consistent.

\subsection{Quantum resources}

In Fig.~\ref{fig:resources}(a) we plot the ensemble-averaged von-Neumann entanglement entropy for a contiguous subsystem $S$ of the chain
\begin{equation}
    \langle S^{\rm vN} \rangle_R = \frac{1}{R} \sum_r S^{\rm vN}(\ket{\psi_r})
\end{equation}
versus $|S|$ at $N=13$. Here, the von-Neumann entanglement entropy of a pure state $\ket{\psi}$ is $S^{\rm vN}(\ket{\psi})=-\text{tr}(\rho^S \text{ln}\rho^S)$ with $\rho^S = \text{tr}_{\bar{S}}\ket{\psi}\bra{\psi}$. We find that on average the variational states appear to be area-law entangled because the values saturate with increasing $|S|$. Regardless of the scaling with $|S|$, we note that the entanglement of the variational  states is much smaller than the average entanglement of eigenstates within the broadened microcanonical window. In particular, in Fig.~\ref{fig:resources}(b) we compute the broadened microcanonical average of the von Neumann entropy
\begin{align}
\label{eq:smc}
\langle S^{\rm vN} \rangle_{\lambda,\delta} = \sum_E \mathcal{D}^{-1}(\lambda) G_\delta (E-\lambda) S^{\rm vN}(\ket{E}).
\end{align}
Comparing Fig.~\ref{fig:resources}(a) and (b), we observe that the variational states have an order of magnitude less entanglement than the eigenstates which they are superpositions of. Panel (b) also contains the Page entropy, i.e., the average entanglement of states drawn Haar-randomly from the full Hilbert space~\cite{page_average_1993}. We find that the $\lambda=0$ microcanonical average of the entanglement entropy is quite close to the Page value, providing an additional check that the MFIM is highly nonintegrable for the chosen parameters. 

The low entanglement of the variational states can be attributed to the fact that they are generated by low-depth circuits, which makes them atypical. In Fig.~\ref{fig:resources}(c) we plot the average number of layers yielding convergence, which is a proxy for the circuit depth needed to prepare the converged variational state (the depth of a $p$-layer ansatz circuit is $3p$, see Fig.~\ref{fig:ansatz}). We show the behavior for $\lambda/N=-0.5$, but we find a linear scaling in $N$ for the other energy densities as well (with slightly different slopes) such that they
require a similar circuit depth. Since the total number of variational parameters is $2Np^*$, the total number of gates scales roughly quadratically in $N$. The actual number of variational parameters at $N=13$ and $\lambda/N=-0.5$ is only about $78$.

\begin{figure}
    \centering\label{fig:tolerance}
    \setlength{\abovecaptionskip}{-5pt}
    \setlength{\belowcaptionskip}{-5pt}\includegraphics{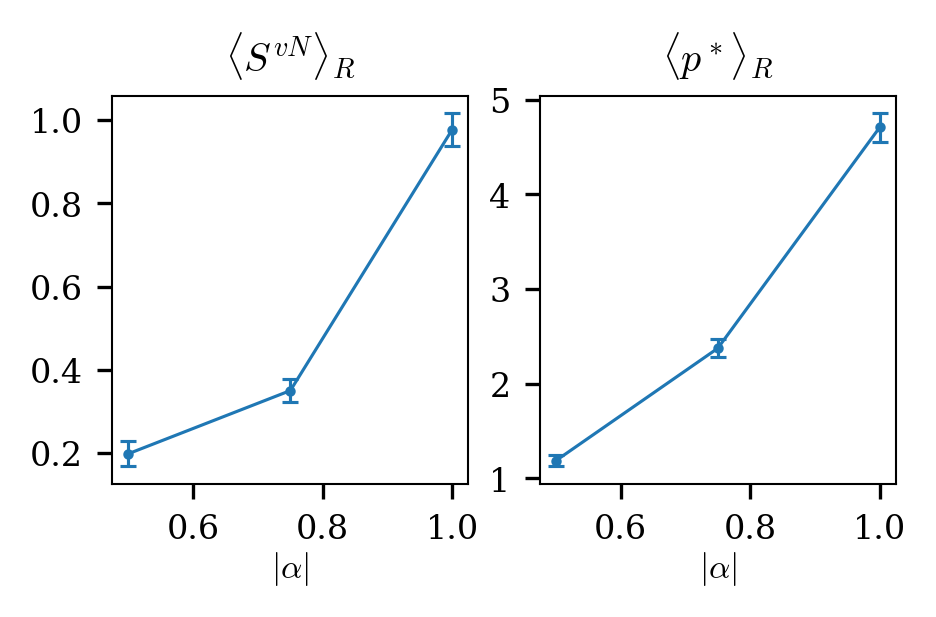}
    \caption{Effect of the tolerance parameter on the half-chain ($|S|=4$) entanglement entropy and circuit depth of the VME, for $N=8$ at target energy density $\lambda/N=-0.75$. The results are averaged over $48$ variational states with error bars being the standard error.}
\end{figure}

We also briefly consider how the entanglement and the number of layers in the variational ansatz depend on the tolerance $\delta$. In particular we consider exponents $\alpha=-0.5,-0.75,-1$ [see Eq.~\eqref{eq:delta}]; the addition of tolerances stricter than $\alpha=-0.5$ used elsewhere in this work limits the numerically accessible system sizes to around $N=8$, similar to Ref.~\cite{zhang_adaptive_2021}. Fig.~\ref{fig:tolerance} shows that both entanglement and circuit depth increase with $|\alpha|$. For example, for $|\alpha|=1$, already at $N=8$ we need $p^*\approx5$ layers for convergence, making this scaling prohibitive for the classical optimizer. It appears that $p^*(N)$ grows much faster with $N$ at $|\alpha|=1$ than it does for $|\alpha|\leq0.75$. The fact that larger entanglement and higher circuit depths would be needed to prepare variational states with smaller energy variance is consistent with other studies, e.g.~Ref.~\cite{banuls_entanglement_2020}.

\section{Conclusion and Outlook}

In this work we propose a VQA for estimating Gaussian microcanonical averages of local operators at intermediate energy density. Given the target average energy $\lambda$ and a microcanonical width of $O(N^{-1/2})$, the variational algorithm evolves random product states into weakly entangled states whose diagonal ensembles are approximately Gaussian-microcanonical on average. We have systematically examined what we call the diagonal and off-diagonal contributions to the error in this estimation, and found that the latter is the dominant source of error. The mean-square off-diagonal error is on the one hand parametrically reduced by ensemble averaging as $R^{-1}$ for small $R$, but on the other hand, for large $R$ saturates to a small value $|c|$ whose size depends principally on the observable under consideration. We have left the identification of the mechanism behind this bias and methods to remove it for future work.


We have also examined the performance of the algorithm in an observable-independent way by computing the trace distance between the subsystem variational ensemble and the subsystem microcanonical one, which appears to continually decrease with $N$ when we take $R = O(N^2)$ and $\lambda/N=-0.75,-0.5$ (though we cannot rule out saturation), whereas for $\lambda/N=-0.25,0$ there is not a consistent decay. We find this result interesting because for other finite temperature VQA methods, intermediate temperatures (as opposed to infinite temperature) are more difficult to simulate \cite{verdon_quantum_2019,selisko_extending_2022}. Since the number of variational parameters appears to scale roughly as $O(N^2)$, this suggests that a classical optimizer could handle the optimization at larger system sizes. The main bottleneck in the classical simulation of our proposed VQA is the repeated evaluation of the cost function; it would be useful to implement a more sophisticated classical simulation of the QPU, for example by tensor network methods so that larger systems could be reached. This would help to decide if the trend in the trace distance continues for larger $N$.

Before concluding, let us briefly discuss the complexity of the VME algorithm. Neglecting the non-zero bias constant $|c|$ on the basis that it is small, we note that the computational time complexity of VME is $O(MRGS)$ where $M$ is the number of times the cost function is requested during the optimization, $R$ is the number of states in the variational ensemble, $G$ is the number of gates in the variational circuit, and $S$ is an upper bound on the number of shots needed to estimate the cost function during each evaluation.

Let $\Omega = (H-\lambda)^2$. In the beginning of the optimization, when the state $\ket{\psi(\bm{\theta})}$ is basically a product state, we have
\begin{equation}
    \text{Var}(\Omega) \sim O(N^2),
\end{equation}
implying that $O(N^2)$ shots are needed to get a system-size independent statistical error. This can then be further reduced to the desired tolerance by an $O(1)$ in system size multiplicative factor. Towards the end of the optimization, we'll end up with a state having
\begin{equation}
    \text{Var}(\Omega) \sim O(N^{-2})
\end{equation}
(neglecting the small non-Gaussian tails of $\rho_R(E)$ that were discussed in Sec.~\ref{subsec:diagonal_error}). Thus, the most ``shot costly" part of the optimization is in the beginning. To simply upper bound the resources, we thus take $S = O(N^2)$.

Now, we have observed empirically that $G = O(N^2)$, and that the off-diagonal error scales as $R^{-1/2}$ (again neglecting $|c|$). Therefore, since the diagonal error is $O(N^{-1})$, we may take $R=O(N^2)$ and conclude that the time complexity of VME is $O(MN^6)$ in achieving a statistical error of $O(N^{-1})$. Whether or not the number of calls $M$ is exponentially large in $N$ is still an open question for all variational quantum algorithms.

In Sec.~\ref{section:diagonal_error} we saw that the diagonal error vanishes as $O(\delta/N)$ with increasing system size, so that even product states [whose typical energy width is $O(\sqrt{N})$] with target energy $\lambda$ would suffice for a vanishing diagonal error, provided the ETH holds. An ensemble of $R$ random product states would also yield an off-diagonal error proportional to $R^{-1/2}$, but it is unclear if one can variationally squeeze those states onto a microcanonical window without violating the unbiased condition $\mathbb{E}[x_r]=0$.



As far as variational algorithms are concerned, the VME could be considered as an example of a broader class of VQAs where the convergence criterion is based on the value of the cost function rather than its gradient. Furthermore, the smallness of the cost function at convergence is only $O[1/{\rm poly} (N)]$ while all initial random product states were able to converge, so it would be interesting to study if the well known barren-plateau problem \cite{mcclean_barren_2018,larocca_diagnosing_2022} is less significant in this setting.

\acknowledgments
The authors acknowledge valuable discussions with Anatoly Dymarsky, Andreas Elben, Ronak Tali, Milan Kornja\v{c}a, Yihua Qiang, Ana-Marija Nedi\'{c}, Niladri Gomes, Yong-Xin Yao, and Laura James. This material is based upon work supported by the National Science Foundation under Grant No.~DMR-2038010.

\appendix{}
\begin{widetext}
\section{Density of states}\label{section:dos}

\begin{figure}
    \centering\label{fig:dos}
    \setlength{\abovecaptionskip}{-5pt}
    \setlength{\belowcaptionskip}{-5pt}
    \includegraphics{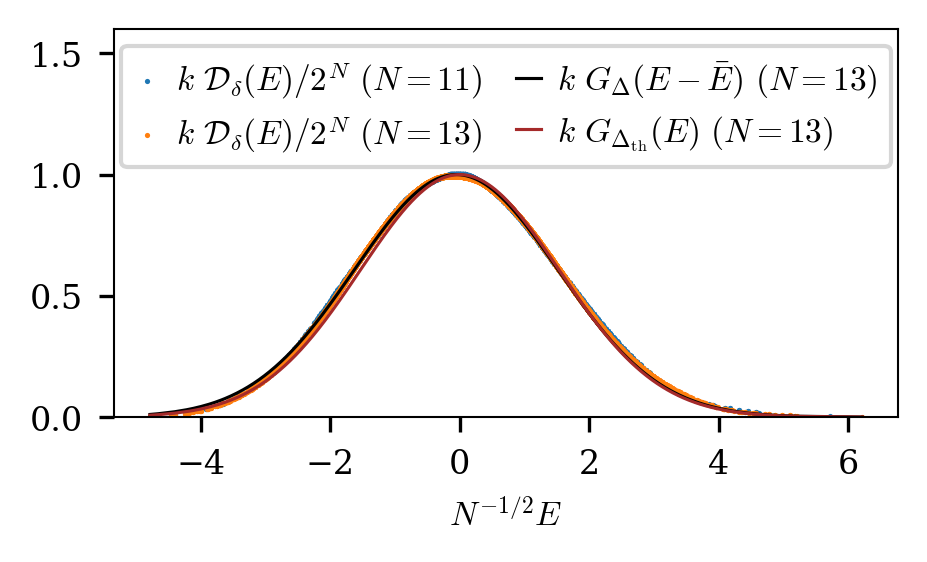}
    \caption{The broadened density of states of the MFIM plotted against $N^{-1/2}E$ so that the form of the curves is $N$-independent. At $N=13$, a Gaussian best-fit yields parameters $\gamma=\Delta^2/N = 2.47$ and $\bar{E}/\Delta=-0.03$ at $N=13$. The theoretical value calculated directly from $H$ with $\gamma_\text{th}=\Delta_\text{th}^2/N=2.35$ is also plotted and seen to agree well. The prefactor $k$ depends on the curve and is chosen so that the maximum value of each curve is $1$ (the factor is $(2\pi\Delta^2)^{1/2}$ when the width is $\Delta$).}
\end{figure}

Let $G_y(x) = (2\pi y^2)^{-1/2}e^{-x^2/2 y^2}$ be a normalized Gaussian window function with zero mean and standard deviation $y$. We maintain the convention that $f(Q) = \sum_q f(q) \ket{q}\bra{q}$ for some Hermitian operator $Q$ with eigenvalues $q$. We find numerically that the broadened density of states $\mathcal{D}_\delta (\lambda)=\text{tr } G_\delta(H-\lambda)$, with broadening parameter $\delta = (\Delta E/N) N^{-1/2}$ is smooth (away from the tails of the spectrum) and can be approximated by a Gaussian of the form $2^N G_\Delta(E) = 2^N (2\pi\Delta^2)^{-1/2}e^{-E^2/2\Delta^2}$ with $\Delta^2 = \gamma N$ and $\gamma$ becoming $N$-independent in the thermodynamic limit. The density of states and a Gaussian best fit curve are shown in Fig.~\ref{fig:dos}. Following the method discussed in the Appendix of \cite{dymarsky_subsystem_2018}, we can check if this approximation is reasonable by estimating the parameter $\gamma$ from the Hamiltonian directly. Note that in terms of the supposed form of the density of states and in the thermodynamic limit the following equality should hold
\begin{equation}
    \text{tr}(H^2) = \int_{-\infty}^\infty \mathrm{d}E\, \mathcal{D}(E) E^2 = 2^N \gamma N.
\end{equation}
With periodic boundary conditions it can be checked for the Hamiltonian~\eqref{eq:mfim} that $\text{tr}(H^2)=2^N [(1+h_z^2) N + \sum_j h^2_{xj}]$, but since the random fields $h_{xj}$ have been chosen to all be within $1\%$ of the central value $h_x$, we can safely approximate $\text{tr}(H^2)\approx2^N (1+h_x^2+h_z^2)N$, yielding the theoretical estimate $\gamma_{\rm th} = 1+h_x^2 + h_z^2=2.35$, which is within about $5\%$ of the best fit value of $\gamma$. We also checked that the density of states is roughly unaffected when it is computed using using the broadening parameter $\sigma$ (with which the variational states converge) instead of the broadening parameter $\delta$.

\section{Off-diagonal contribution assuming independent identically distributed random variables}\label{section:iid}

In Section.~\ref{section:motivation}, we claimed that should the order-one fluctuations $R_{EE'}$ be actual independent and identically distributed random variables, then the off-diagonal contribution to equation Eq.~\eqref{eq:psi_A_psi} would be typically $O(\mathcal{D}^{-1/2}(\lambda))$. To see this, assume for $E>E'$ that $R_{EE'}$ are samples from an underlying distribution satisfying $\mathbb{E}[R_{EE'}]=0$, $\mathbb{E}[R_{EE'}^2]=1$, and $\mathbb{E}[R_{EE'}R_{E''E'''}]=0$ for $E\neq E''$, $E'\neq E'''$ and $E''>E'''$. We restrict the energies to the upper triangular part of the $R$ matrix since $R_{EE'}=R_{E'E}$ for energy-basis real observables. Let
\begin{equation}
    x_1 = 2 \sum_{E>E'\in W} c_E c_{E'} \braket{E|A|E'} = 2 \sum_{E>E'\in W} c_E c_{E'} \mathcal{D}^{-1/2}(\bar{E}) f(\bar{E},\omega) R_{EE'},
\end{equation} where in the second equality we have inserted the ETH matrix element ansatz. The above is notation consistent with Section.~\ref{section:micro-sup}, where $x_1$ is the ``off-diagonal error" for a single microcanonical superposition state supported only on the microcanonical window $W$. Clearly we have $\mathbb{E}[x_1]=0$ and
\begin{equation}
    \mathbb{E}[x_1^2] = 4 \sum_{\substack{E > E' \in W \\ E'' > E'''\in W}} c_E c_{E'} c_{E''} c_{E'''} \mathcal{D}^{-1/2}(\bar{E}) \mathcal{D}^{-1/2}(\bar{E''}) f(\bar{E},\omega) f(\bar{E''},\omega'') \mathbb{E}[R_{EE'} R_{E''E'''}]
\end{equation}
where $\bar{E'} = (E''+E''')/2$ and $\omega'' = E''-E'''$. The independence assumption collapses the quadruple sum giving
\begin{equation}\label{eq:iid_result}
    \mathbb{E}[x_1^2] = 4 \sum_{E > E' \in W} c^2_E c^2_{E'}\mathcal{D}^{-1}(\bar{E}) f^2(\bar{E},\omega).
\end{equation}
Normalization of the state implies $c^2_E= O(1/n)$ where $n$ is the number of eigenenergies in $W$, and the double sum runs over $n(n-1) = O(n^2)$ energies. Altogether we have $\mathbb{E}[x_1^2] = O(\mathcal{D}^{-1}(\lambda))$ since $\lambda$ is a typical energy in the window, and where we have neglected $N$ dependence of $f(\bar{E},\omega)$; see the footnote below Eq.\eqref{eq:explicit_diagonal_error}. In the thermodynamic limit then, the central limit theorem implies that $x_1 \approx O(\mathcal{D}^{-1/2}(\lambda))$.

\section{Expectation value of $(H-\lambda)$ in broadened microcanonical ensemble}\label{section:broadened}

Here we justify Eq.~\eqref{eq:smooth_eth}, which expresses the relation between the smooth function $A(E)$ appearing in the ETH matrix element ansatz and the microcanonical expectation value $\langle A \rangle_{\lambda,\delta}$,  by considering the expectation value of $(H-\lambda)$ in the smooth microcanonical ensemble. We take the density of states to be the Gaussian function described in Appendix~\ref{section:dos}. Under such an assumption, the first few terms of the Taylor series for the density of states $\mathcal{D}(E)=2^N (2\pi\Delta^2)^{-1/2}e^{-E^2/2\Delta^2}$ near energy $\lambda$ read
\begin{equation}\label{eq:dos_expand}
    \mathcal{D}(E)/\mathcal{D}(\lambda) = 1 - \frac{\lambda}{\Delta^2}(E-\lambda)
    + \left[\left(\frac{\lambda}{\Delta^2}\right)^2 - \frac{1}{\Delta^2}\right]\frac{(E-\lambda)^2}{2}
    + \frac{1}{\Delta^4}\left(3\lambda-\frac{\lambda^3}{\Delta^2}\right)\frac{(E-\lambda)^3}{6} + \cdots.
\end{equation}
Here, $\Delta^2 = \gamma N$ with $\gamma \approx 1 + h_x^2 + h_z^2$, and we have not assumed that $E-\lambda$ is small in any sense yet, only that the density of states admits a Taylor expansion in the thermodynamic limit. If we ignore any error incurred in replacing sums by integrals in the thermodynamic limit, where the energy bandwidth approaches infinity and the level spacing zero, the first moment of $(H-\lambda)$ in the broadened microcanonical ensemble
$\rho_{\lambda,\delta}= \sum_E \mathcal{D}^{-1}(\lambda) G_\delta(E-\lambda) \ket{E}\bra{E}$ reads
\begin{align}\label{eq:moments_expand}
    &\text{tr}[(H-\lambda) \rho_{\lambda,\delta}] = - \frac{\delta^2\lambda}{\Delta^2} + \frac{3\delta^4}{6\Delta^4}\left(3\lambda-\frac{\lambda^3}{\Delta^2}\right) + \cdots
\end{align}

We can then use these formulae to estimate the expectation value of an ETH-obeying operator in this broadened microcanonical ensemble. In doing so, an important consequence of the ETH is that the smooth function $A(E)$ should be expressible as a function of energy density in the thermodynamic limit, see Fig.~\ref{fig:eth} in the main text. In this paper we consider only Pauli-string-type observables. In this case, note that $A(E) = a(E/N)$ is $O(1)$ because $A(E)$ is defined via (a best fit curve to) the averaging procedure
\begin{equation}
    \frac{1}{K}\sum_{E'} \braket{E'|A|E'}
\end{equation}
over $K$ eigenstates near $\ket{E}$, and $|\braket{E'| A |E'}|\leq 1$ for Pauli strings. Now since $A(E) = a(E/N)$, it follows that $a(x) = O(1)$ and 
\begin{equation}
    \frac{\mathrm{d}A}{\mathrm{d}E}\bigg|_E = \frac{1}{N}\frac{\mathrm{d}a}{\mathrm{d}x}\bigg|_{E/N} = O\bigg(\frac{1}{N}\bigg),
\end{equation} 
and similarly for higher derivatives. Now consider a broadened microcanonical ensemble at energy $\lambda$, i.e. $\rho_{\lambda,\delta} = \mathcal{D}^{-1}(\lambda)G_\delta(H-\lambda)$. Then, the expectation value of the \emph{smooth ETH function} in this ensemble is obtained by going to the continuum and combining Eqs.~\eqref{eq:dos_expand} and \eqref{eq:moments_expand}. The result reads 
\begin{equation}
    \int \mathrm{d}E \frac{\mathcal{D}(E)}{\mathcal{D}(\lambda)} G_\delta(E-\lambda) A(E) = A(\lambda) + O(\delta^2/N).
\end{equation}
From this and the ETH ansatz follows Eq.~\eqref{eq:smooth_eth}.

\section{Additional discussion of diagonal ensemble}\label{section:var_ens}

In this Appendix we explain the deviations in $\mu$ and $\sigma$ from $\lambda$ and $\delta$, respectively, when fitting the converged variational ensembles $\rho_R$ to the best fit curves $\rho_{\mu,\sigma}$. As was described in Sec.~\ref{section:diagonal_error} of the main text, $\mu$ will under-shoot $\lambda$ because the density of states is non-uniform. We can confirm this more precisely as follows. Using the best fit parameters to the variational ensemble (i.e. the values in Table \ref{tab:converged_ensembles}), we treat the spectrum as continuous and compute the numerical integral $\text{tr}[\rho_{\mu,\sigma}(H-\lambda)]\approx -0.042$ at $\lambda/N=-0.5$. Doing the same for the ensemble whose central energy is $\lambda$, we find that the value of $\text{tr}[\rho_{\lambda,\delta}(H-\lambda)]\approx0.142$. In the latter calculation, we emphasize that this value is roughly the same whether using $\delta$ or $\sigma$ for the width of the Gaussian. Thus, the ensemble with central energy $\mu$ actually minimizes the operator $(H-\lambda)$ much better than the ensemble with central energy $\lambda$. Thus the deviation in $\mu$ from $\lambda$ is a finite-size effect due to a non-uniform density of states.


We now address the deviations in $\sigma$ from $\delta$, which we claim to be due to the slight ``non-Gaussianity" of $\overline{\rho_R}(E)$, i.e. the excess weight outside the Gaussian window that we see in Fig.~\ref{fig:diagonal_error}. Best-fit curves aside, we first check that the fluctuations of $\rho_R(E)$ around $\overline{\rho_R}(E)$ contribute negligibly to the expectation value of $(H-\lambda)^2$. We directly compute $\text{tr}[\overline{\rho_R}(H)(H-\lambda)^2]\approx 0.92\, \delta^2$, and we can see that this value is consistent with the actual ensemble average value of the cost function, $\text{tr}[\rho_R(H-\lambda)^2]\approx 0.90\, \delta^2$. The fact that these values are slightly less than $\delta^2$ can be attributed to the convergence criterion only requiring that the variance (which is approximately the cost) be \textit{at most} $\delta^2$. Now comparing this to the Gaussian model best-fit ensemble,which  predicts a cost function value of only $\text{tr}[\rho_{\mu,\sigma}(H-\lambda)^2]\approx 0.69 \delta^2$, we see that it undershoots $\delta^2$ since it is missing the contribution from the excess energy weight.


\section{Justification for replacing $\hat{A}$ by $\tilde{A}$ and further properties of $\tilde{A}$}\label{section:truncate}

\begin{figure}
    \centering\label{fig:sub_matrix}
    \setlength{\abovecaptionskip}{-5pt}
    \setlength{\belowcaptionskip}{-5pt}\includegraphics[width=.45\textwidth]{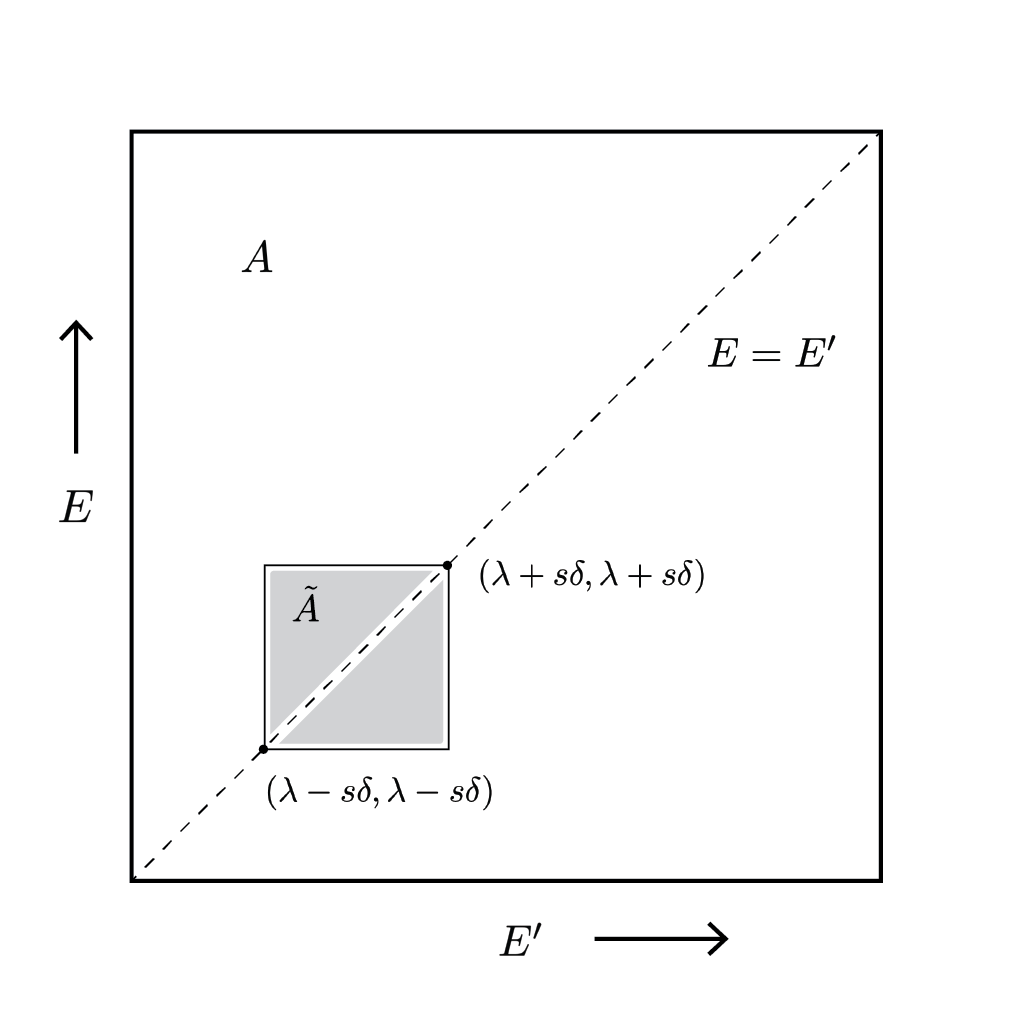}
    \caption{In shaded gray, the off-diagonal only sub-matrix $\tilde{A}$ of $A$ relevant to off-diagonal error for microcanonical superpositions whose weight is mostly within $s$ standard deviations $\delta$ of the central energy $\lambda$. Matrix elements are shown on the energy scale rather than the eigenvalue index scale.}
\end{figure}

\begin{figure}
    \centering\label{fig:truncate}
    \setlength{\belowcaptionskip}{-5pt}
    \setlength{\abovecaptionskip}{-5pt}
\includegraphics{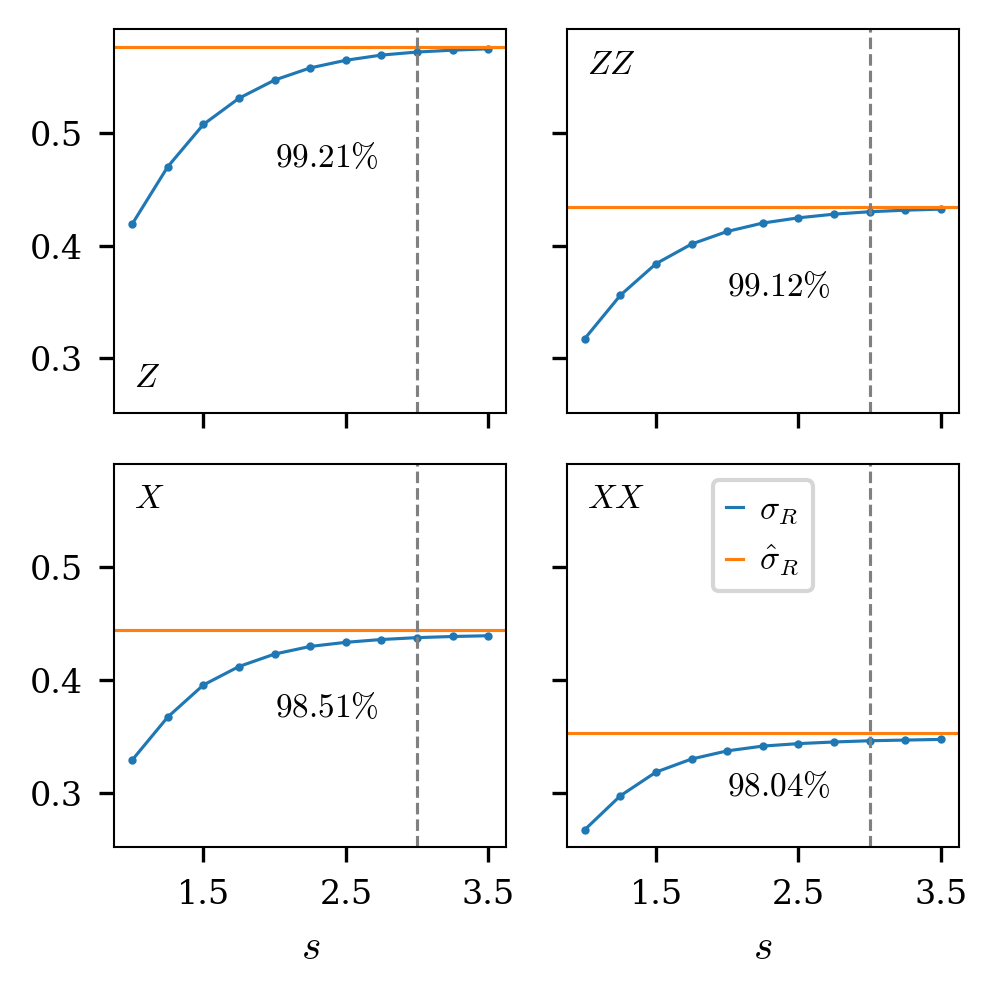}
    \caption{For various operators $A$, the effect of expanding the window to include $s$ standard deviations around $\lambda$ for $N=13$, $R=288$, and when $\lambda/N=-0.5$. We see that at $s=3$, we have captured basically all of $\hat{\sigma}_R$ as shown by the percentages. Other energy densities are unremarkable except that at $\lambda/N=-0.75$ only about $97\%$ of $XX$ is captured.}
\end{figure}

\begin{figure}
    \centering
    \setlength{\abovecaptionskip}{-5pt}
    \setlength{\belowcaptionskip}{-5pt}\includegraphics{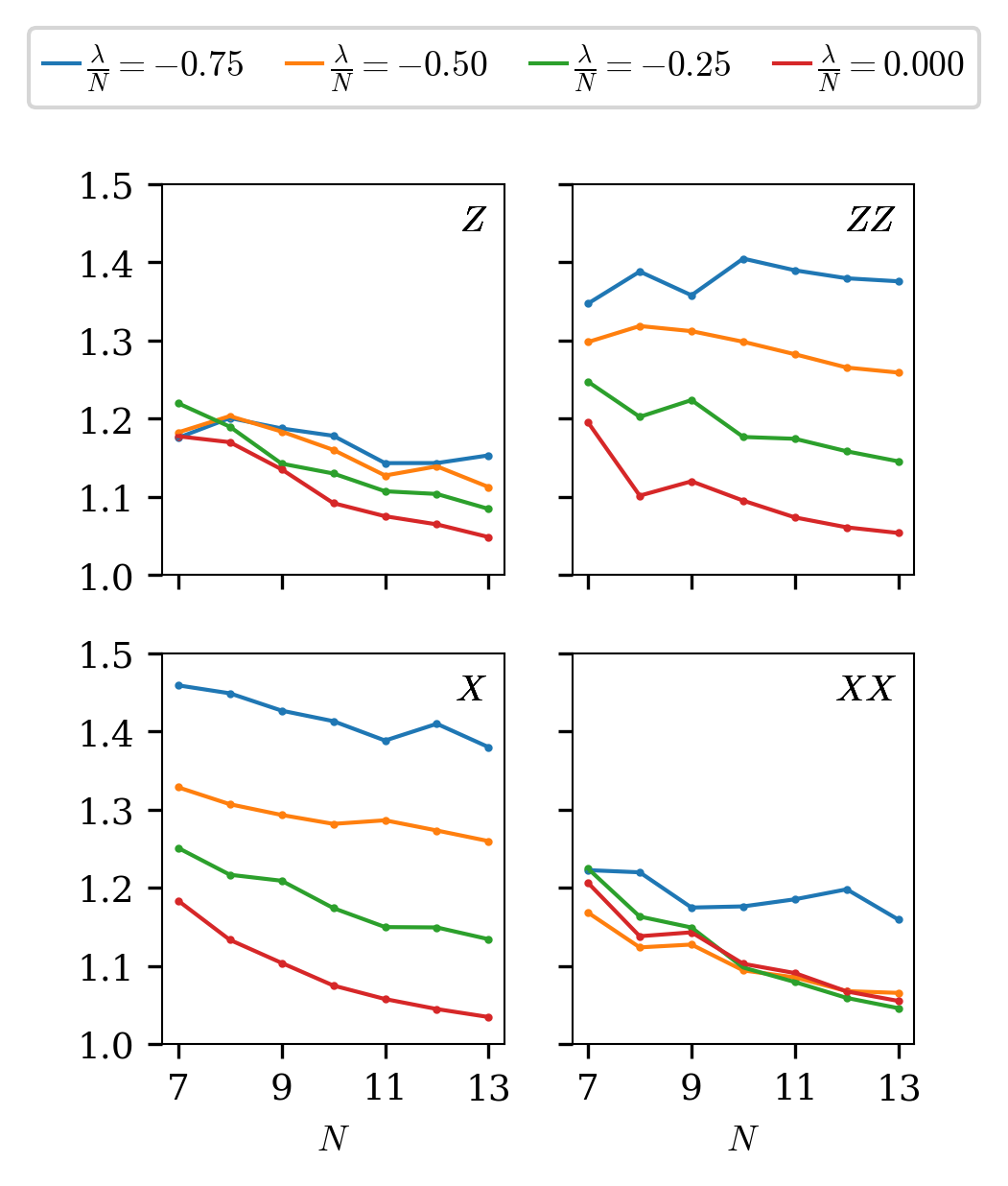}\label{fig:sing_val}
    \caption{At various energy densities $\lambda/N$, the maximum singular value of $\tilde{A}$ for various $A$, with $\tilde{A}$ the $3\delta$ large sub-matrix of $A$ and $\delta=O(N^{-1/2})$, i.e. with $s=3$ as in Fig.~\ref{fig:sub_matrix}.}
\end{figure}

\begin{figure}
    \centering
    \setlength{\abovecaptionskip}{-5pt}
    \setlength{\belowcaptionskip}{-5pt}\includegraphics{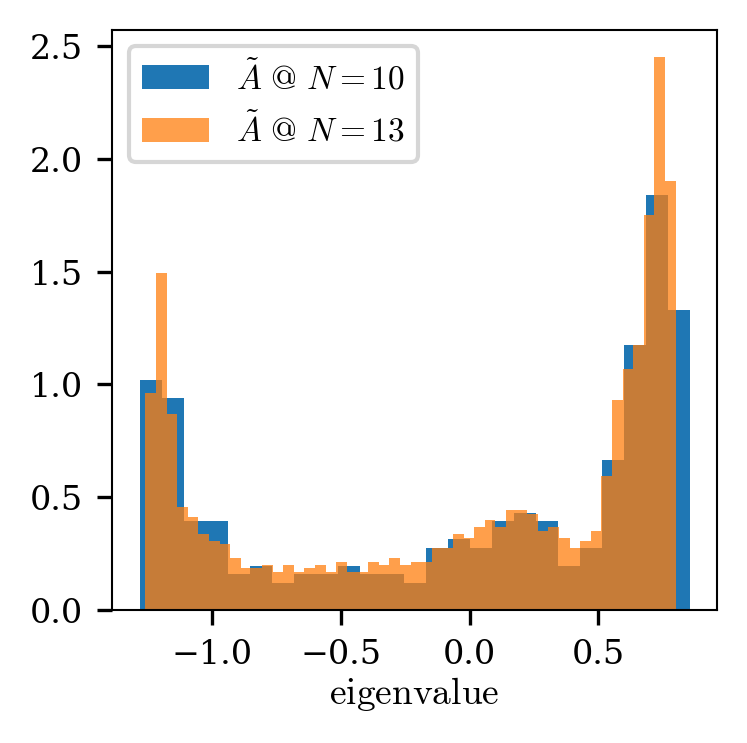}\label{fig:rmt_N_depend}
    \caption{For the operator $A=X$ at energy density $\lambda/N=-0.5$, probability density functions of the eigenvalues of $\tilde{A}$ as in the orange histogram of Fig.~\ref{fig:off_diagonal_error}(c) in the main text, but here shown for $N=10$ in blue and for $N=13$ in transparent orange.}
\end{figure}

In this Appendix we justify deleting certain energy basis matrix elements of $A$ based on the form of the variational states, for the purposes of gaining some intuition about the nature of the off-diagonal error. Let the operator $\hat{A}$ be $A$ with its energy basis diagonal elements set to zero, i.e.,
\begin{align}
    \braket{E|\hat{A}|E'} =
    \begin{cases}
    \braket{E|A|E'} \ &\text{if}\ E\neq E' \\
    \quad\ \  0 \quad &\text{if}\ E = E'.
    \end{cases}
\end{align}
At this point we also define $\tilde{A}$ for general $s$, where $s$ is the number of standard deviations $\delta$ around $\lambda$ that are $\textit{not}$ deleted:
\begin{align}
    \braket{E|\tilde{A}|E'} =
    \begin{cases}
    \braket{E|A|E'} \ &\text{if $E\neq E'$, $|E-\lambda|\leq s\delta$,}\\
                      &\text{and $|E'-\lambda|\leq s \delta$}\\
    \quad\ \  0 \quad &\text{otherwise}.
    \end{cases}
\end{align}
An equivalent definition is given graphically in Fig.~\ref{fig:sub_matrix}. Since the ensemble of variational states $\ket{\psi_r}$ approximates a Gaussian microcanonical ensemble near with average energy near $\lambda$ \textit{on average} [see Fig.~\ref{fig:diagonal_error}(a)], we can anticipate that the value of $\braket{\psi_r|\hat{A}|\psi_r}^2$ will also be unaffected \textit{on average} by replacing $\hat{A}$ with $\tilde{A}$ when $\tilde{A}$ has a sufficiently large energy support. Because of the established Gaussian form, we might expect that two standard deviations around $\lambda$ is always sufficient to capture $95\%$ of the average absolute error. However there are fluctuations around this behavior, and the coarse grained variational states have some excess energy weight beyond the Gaussian best-fit curve. Furthermore, the variational ensembles are not exactly centered on $\lambda$. Thus, we justify replacing $\hat{A}$ with an appropriately chosen $\tilde{A}$ numerically as follows. In Fig.~\ref{fig:truncate} we compare $\sigma_R$ when computed on a window of size $2s\times 2s$ (see Fig.~\ref{fig:sub_matrix} for clarification) and $\hat{\sigma}_R$, which is computed from the entire spectrum. We show on the plots the fraction of $\hat{\sigma}_R$ that is captured by $\sigma_R$ at $s=3$, i.e. roughly three standard deviations, which we consider to be sufficiently large to capture basically all of $\hat{\sigma}_R$.


For the $3\delta$ truncated operators $\tilde{A}$ we have just discussed, in this Appendix we also consider how the maximum singular value of $\tilde{A}$, i.e. the larger of $|\lambda_{\rm max}(\tilde{A})|,|\lambda_{\rm min}(\tilde{A})|$ scales with $N$. The results are shown in Fig.~\ref{fig:sing_val}. These results provide evidence that the numerical prefactor $\sigma_R$ appearing in the statistical description of the off-diagonal error is $O(1)$ in system size. In this Appendix we also consider how the eigenvalue statistics qualitatively vary with $N$, with an example shown in Fig.~\ref{fig:rmt_N_depend} demonstrating that the distribution is qualitatively independent of $N$, except for the slight decrease of the distribution's width with $N$ as reflected in Fig.~\ref{fig:sing_val}.

\section{Additional numerical data for diagonal error}\label{section:diagonal_data}
\begin{figure}[h]
    \centering
    \setlength{\abovecaptionskip}{5pt}    \includegraphics[width=.95\textwidth]{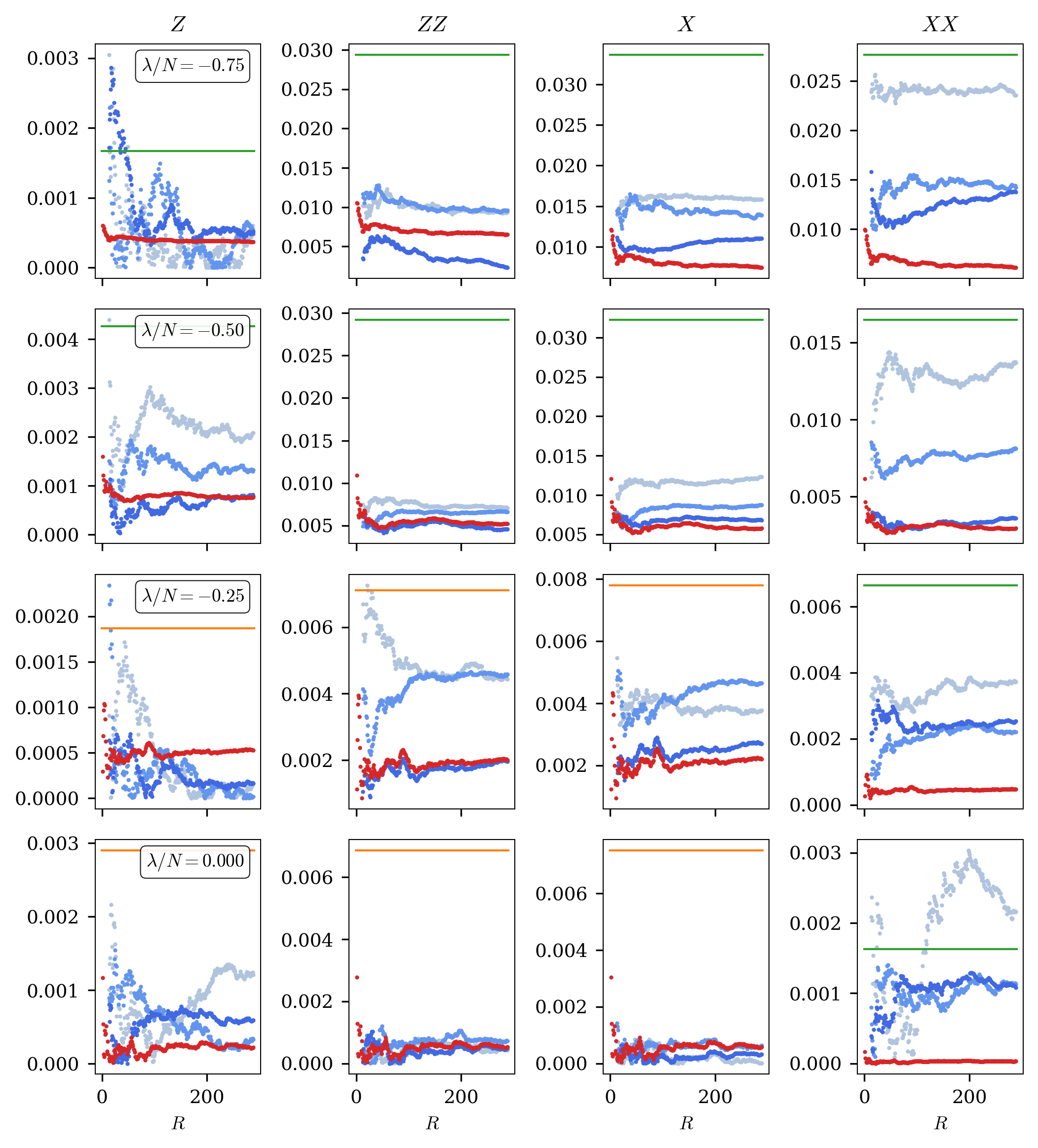}
    \caption{Diagonal error in the VME. Curves are interpreted identically to those in Fig.~\ref{fig:diagonal_error}(b) in the main text, except to bring the estimate $(\delta/N)|a'(\lambda/N)|$ down to scale we plot instead in some cases $(\delta/4N)|a'(\lambda/N)|$. The latter are shown in orange instead of green to indicate the use of a constant scale factor.} 
    \label{fig:more_diag_error}
\end{figure}

Fig.~\ref{fig:more_diag_error} in this Appendix shows the diagonal error in the VME estimate for operators $Z,ZZ,X,XX$ acting in the middle of the chain. As stated in the main text, the results for energy density $\lambda/N=-0.5$ appear to agree best with the prediction of ETH that the error should decay as $1/N$ and agree with $\chi_R/N$ [see Eq.~\eqref{eq:chi_R}] for large $R$. While in general such a clear scaling with $N$ is missing, all cases show that the $N=13$ diagonal error is never larger than some order one fraction of the rough estimate $a'(\lambda/N)\delta/N$, with $XX$ at $\lambda/N=0$ showing the case where the diagonal error comes closest to the estimate. We also note that the ETH prediction $\chi_R/N$ for the diagonal error is always on the correct scale of the error for $N=13$. The operator $XX$ is also an outlier in this sense–$\chi_R/N$ significantly underestimates the actual error except for $\lambda/N=-0.5$.

\clearpage
\newpage

\section{Additional numerical data for off-diagonal error}\label{section:off_diagonal_data}

\begin{figure}[h]
    \centering
    \setlength{\abovecaptionskip}{-5pt}
    \includegraphics[]{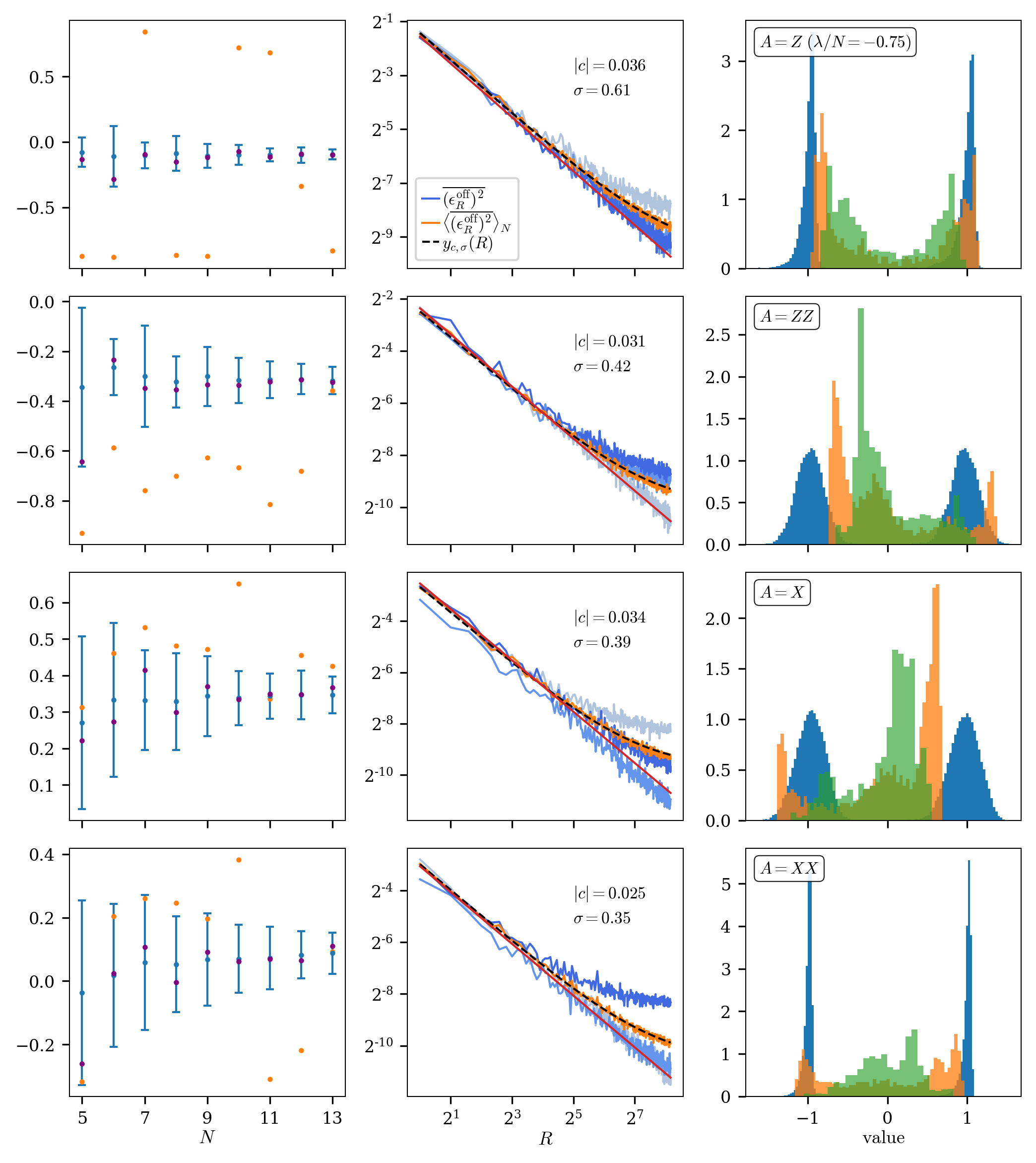}\label{fig:off_diagonal_error_-0.75}
    \caption{Off-diagonal error in the VME for observables $A=Z,ZZ,X,XX$ at $\lambda/N=-0.75$. See the caption of Fig.~\ref{fig:off_diagonal_error} in the main text for further explanation of what is shown in the plots.}
\end{figure}
\begin{figure}[h]
    \centering
    \setlength{\abovecaptionskip}{-5pt}
    \includegraphics[]{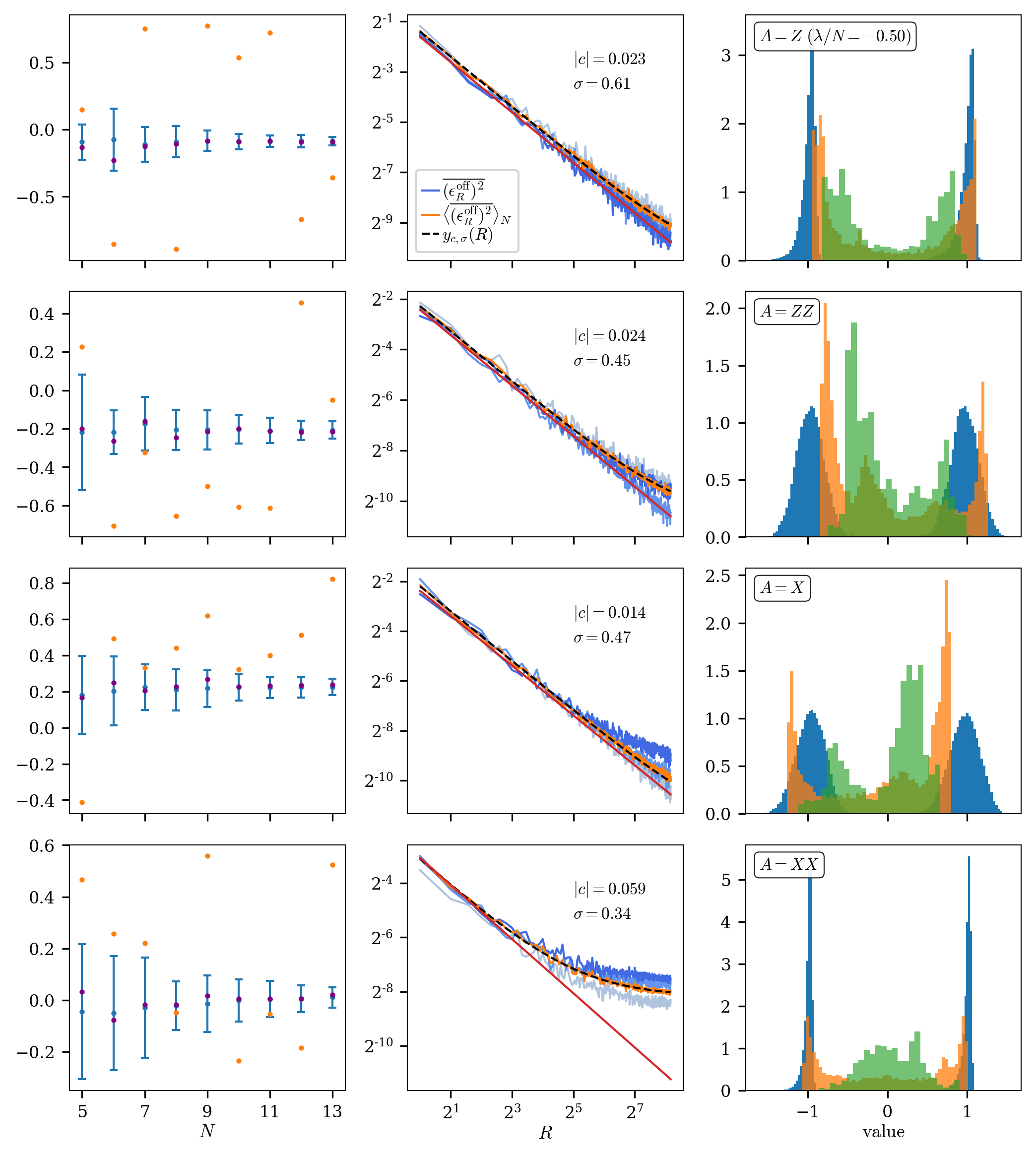}\label{fig:off_diagonal_error_-0.5}
    \caption{Off-diagonal error in the VME for observables $A=Z,ZZ,X,XX$ at $\lambda/N=-0.5$. See the caption of Fig.~\ref{fig:off_diagonal_error} in the main text for further explanation of what is shown in the plots.}
\end{figure}
\begin{figure}[h]
    \centering
    \setlength{\abovecaptionskip}{-5pt}
    \includegraphics[]{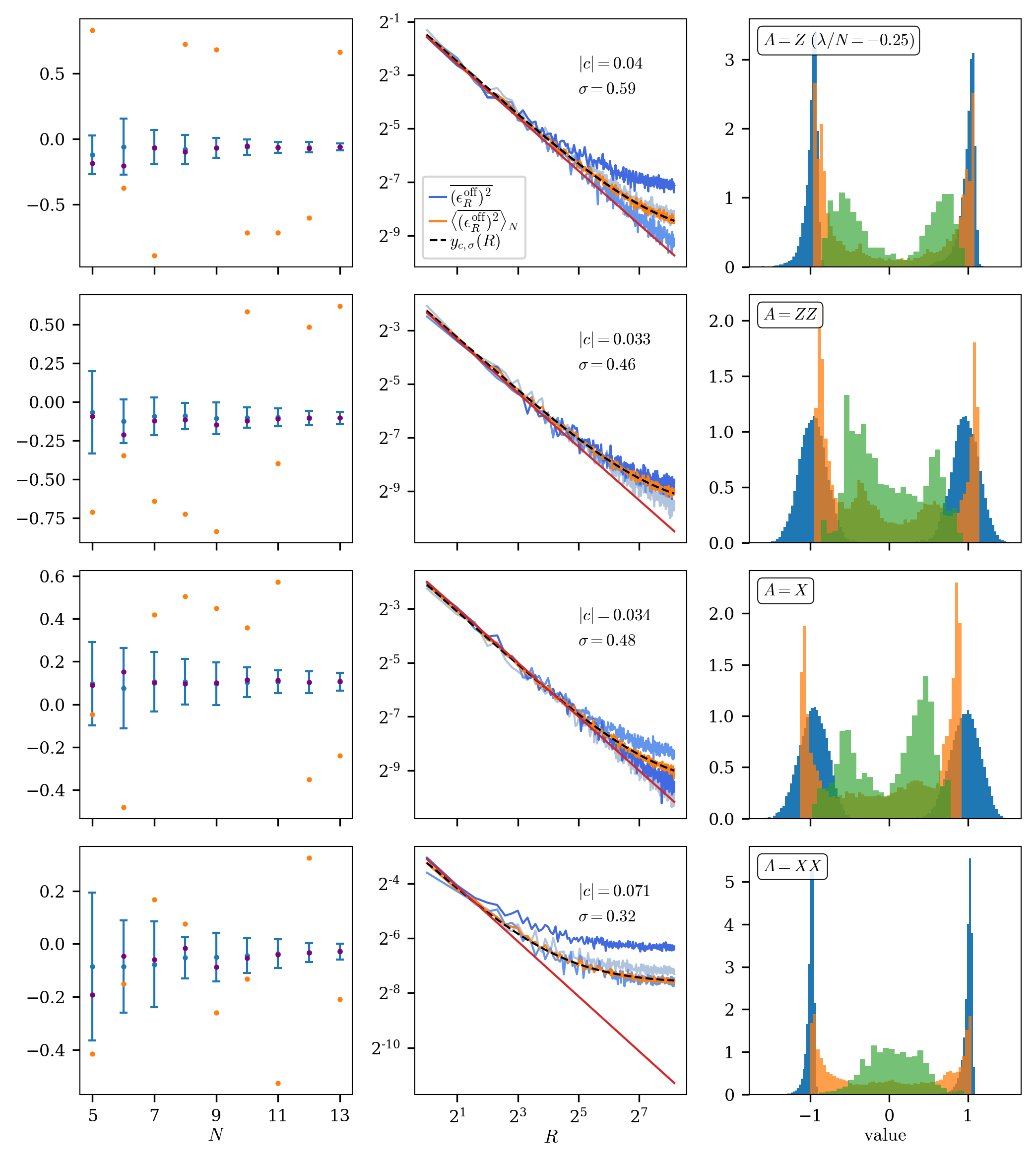}\label{fig:off_diagonal_error_-0.25}
    \caption{Off-diagonal error in the VME for observables $A=Z,ZZ,X,XX$ at $\lambda/N=-0.25$. See the caption of Fig.~\ref{fig:off_diagonal_error} in the main text for further explanation of what is shown in the plots.}
    \label{fig:more_off_diag_0.25}
\end{figure}
\begin{figure}[h]
    \centering
    \setlength{\abovecaptionskip}{-5pt}
    \includegraphics[]{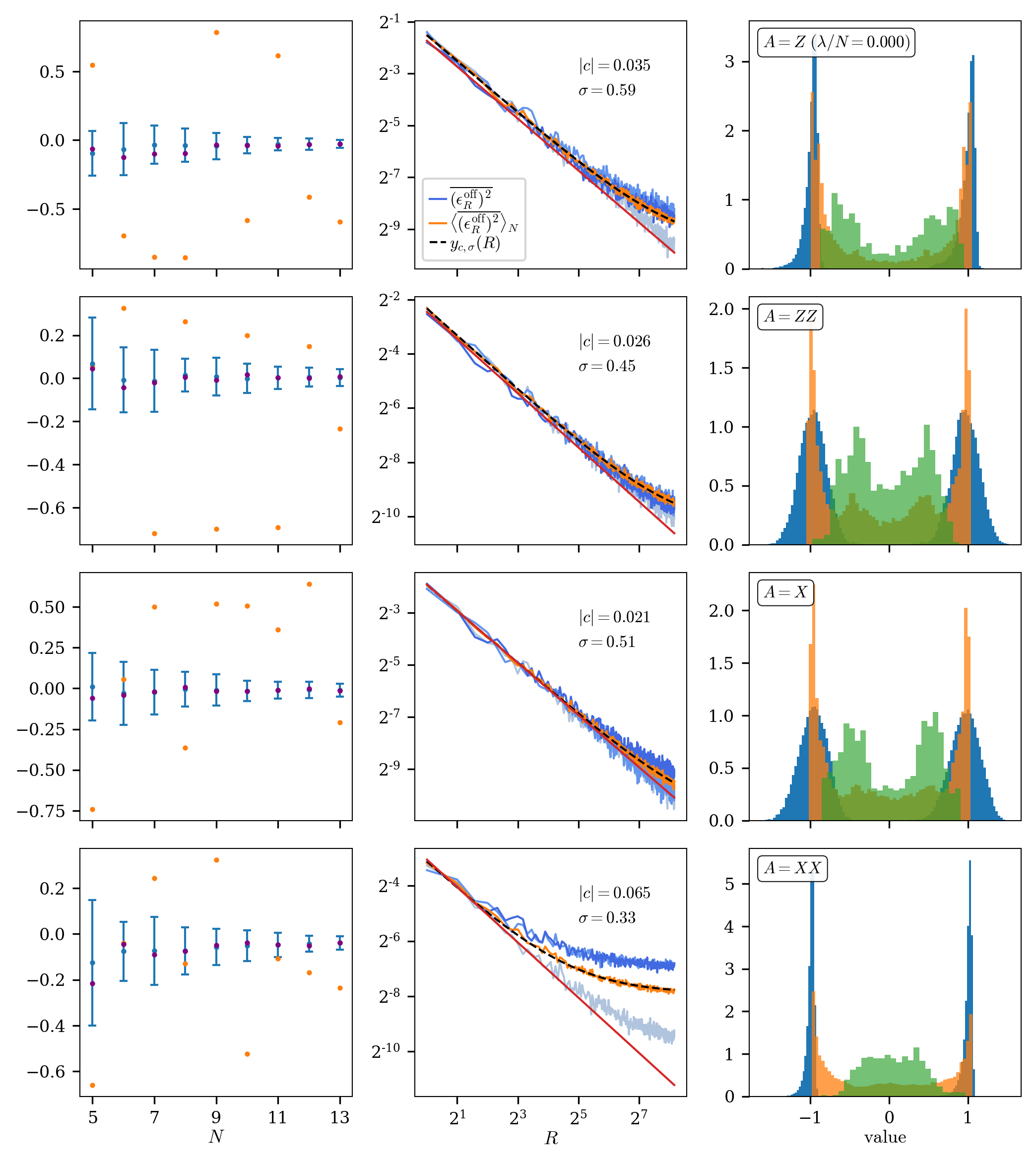}\label{fig:off_diagonal_error_0.0}
    \caption{Off-diagonal error in the VME for observables $A=Z,ZZ,X,XX$ at $\lambda/N=0.0$. See the caption of Fig.~\ref{fig:off_diagonal_error} in the main text for further explanation of what is shown in the plots.}
\end{figure}

In Sec.~\ref{section:numerical} of the main text, Fig.~\ref{fig:off_diagonal_error} demonstrates the behavior of the off-diagonal error for the operator $X$ acting on the middle of the chain at the energy density $\lambda/N=-0.5$. The purpose of this Appendix is to establish that the trends observed there hold more generally across different operators and target energy densities. The plots shown also demonstrate that the operator $XX$ systematically differs from the other considered observables. We show the equivalent of Fig.~\ref{fig:off_diagonal_error} for $\lambda/N=-0.75,-0.5,-0.25,0$ and $A=Z,ZZ,X,XX$ acting on the central one or two sites of the chain. The results are shown in Figs.~\ref{fig:off_diagonal_error_-0.75},\ref{fig:off_diagonal_error_-0.5},\ref{fig:off_diagonal_error_-0.25}, and \ref{fig:off_diagonal_error_0.0}, respectively.

\section{Additional numerical data for expectation values}\label{section:micro_data}

\begin{figure}[h]
    \centering
    \setlength{\abovecaptionskip}{-5pt}
    \includegraphics[width=.95\textwidth]{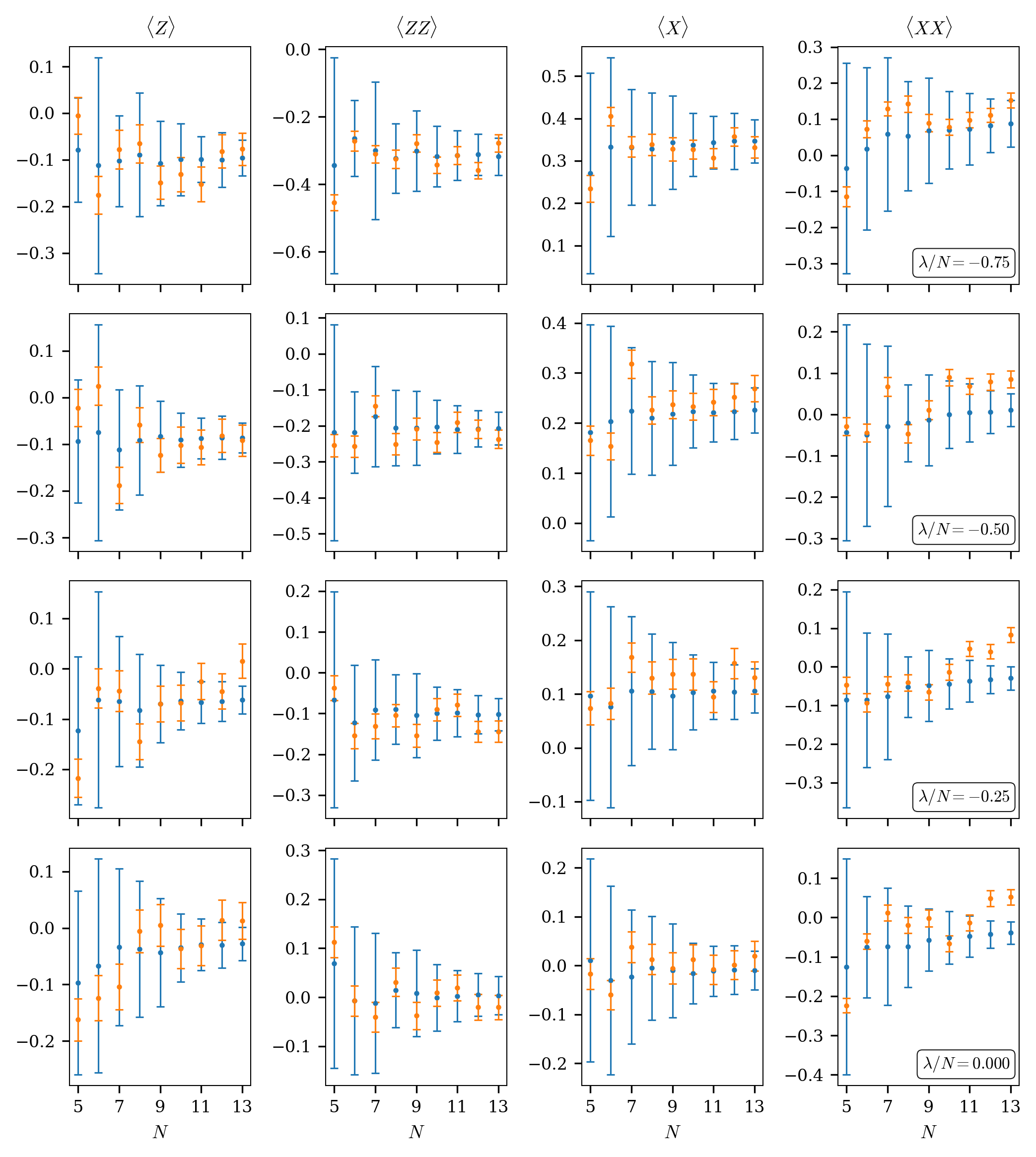}
    \caption{Ensemble-averaged VME observable expectation values (orange) and broadened microcanonical averages (blue) plotted versus system size for the full range of energy densities and operators considered in this work. See the caption of Fig.~\ref{fig:observables} in the main text for further explanation of what is shown in the plots.}
    \label{fig:all_mc_avgs}
\end{figure}

Here we show in Fig.~\ref{fig:all_mc_avgs} the VME estimates for the four local observables $A=Z,ZZ,X,XX$ acting on the central one or two sites of the chain. We show these estimates when targeting four different energy densities $\lambda/N=-0.75,-0.5,-0.25,0$ and when $R=288$. As we observed in Sec.~\ref{section:numerical}, the off-diagonal error generally does not systematically depend on $N$. Here we can see that for fixed $R$, beyond $N\sim 9$, the accuracy of the VME estimates indeed does not depend systematically on $N$ except again for $XX$ where it appears to increase with $N$, consistent with the $R=288$, $N=13$ value of the off-diagonal error for $XX$ being largest in the plots in Appendix.~\ref{section:off_diagonal_data}. We see that the microcanonical estimates can be quite good, as for $Z$ at $\lambda/N=-0.5$, or quite poor, as for $XX$ at $\lambda/N=0$. The better results are generally for the operators $Z,ZZ$, and $X$ and at the lower target energy densities $\lambda/N = -0.75$ and $\lambda/N = -0.5$.

\end{widetext}
\bibliography{references.bib}
\end{document}